\documentclass[fleqn,usenatbib]{mnras}

\usepackage{newtxtext,newtxmath}

\usepackage[T1]{fontenc}

\DeclareRobustCommand{\VAN}[3]{#2}
\let\VANthebibliography\thebibliography
\def\thebibliography{\DeclareRobustCommand{\VAN}[3]{##3}\VANthebibliography}

\usepackage{graphicx}	
\usepackage{amsmath}	
\usepackage{tikz}       
\usepackage{xcolor}
\usepackage{hyperref}

\usepackage{listings}               
\definecolor{codegreen}{rgb}{0,0.6,0}
\definecolor{codegray}{rgb}{0.5,0.5,0.5}
\definecolor{codepurple}{rgb}{0.58,0,0.82}
\definecolor{backcolour}{rgb}{0.95,0.95,0.92}

\lstdefinestyle{mystyle}{
    backgroundcolor=\color{backcolour},   
    commentstyle=\color{codegreen},
    keywordstyle=\color{codepurple},
    numberstyle=\tiny\color{codegray},
    stringstyle=\color{blue},
    basicstyle=\ttfamily\footnotesize,
    breakatwhitespace=false,         
    breaklines=true,                 
    captionpos=b,                    
    keepspaces=true,                 
    numbers=left,                    
    numbersep=5pt,                  
    showspaces=false,                
    showstringspaces=false,
    showtabs=false,                  
    tabsize=2
}
\AtBeginDocument{%
  \DeclareRobustCommand{\J18}{\hyperlink{cite.jerkstrand2018emission}{J18}}
  \DeclareRobustCommand{\Jerk20}{\hyperlink{cite.jerkstrand2020properties}{J20}}
  \DeclareRobustCommand{\vBPI23}{\hyperlink{cite.vanbaal2023modelling}{vB23}}
  \DeclareRobustCommand{\vBPII24}{\hyperlink{cite.vanbaal2024diagnostics}{vB24}}
  \DeclareRobustCommand{\vBPIII26}{\hyperlink{cite.vanbaal2026emission}{vB26}}
}  
\makeatother
\newcommand\kms[1]{{$#1\,\text{km}\,\text{s}^{-1}$}}

\newcommand\txtred[1]{{\color{black}#1}}  
\newcommand{\angstrom}{\mbox{\normalfont\AA}}
\definecolor{lime}{HTML}{A6CE39}
\DeclareRobustCommand{\orcidicon}{%
    \begin{tikzpicture}
    \draw[lime, fill=lime] (0,0) 
    circle [radius=0.16] 
    node[white] {{\fontfamily{qag}\selectfont \tiny ID}};
    \draw[white, fill=white] (-0.0625,0.095) 
    circle [radius=0.007];
    \end{tikzpicture}
    \hspace{-2mm}
}
\newcommand{\orchidJerkstrand}{\href{https://orcid.org/0000-0001-8005-4030}{\orcidicon}}
\newcommand{\orchidVanBaal}{\href{https://orcid.org/0009-0001-3767-942X}{\orcidicon}}

\title[EXplosive TRAnsient Spectral Simulator]{\texttt{ExTraSS}: a Domain Decomposed 3D NLTE Radiative Transfer spectral synthesis code for nebular phase transients}  

\author[van Baal \& Jerkstrand]{
Bart F. A. van Baal\orchidVanBaal,$^{1}$\thanks{E-mail: barteld.vbaal@astro.su.se}
and Anders Jerkstrand\orchidJerkstrand$^{1}$
\\
$^{1}$The Oskar Klein Centre, Department of Astronomy, Stockholm University, AlbaNova, Se-10691 Stockholm, Sweden\\
}

\date{Accepted XXX. Received YYY; in original form ZZZ}

\pubyear{\the\year{}}

\begin{document}
\label{firstpage}
\pagerange{\pageref{firstpage}--\pageref{lastpage}}
\maketitle

\begin{abstract}
In the nebular phase, supernovae are powered by radioactive decay and continuously fade, while their densities have decreased enough such that the expanding nebula becomes (largely) optically thin and the entire structure contributes to the emission. Models for the nebular phase need to take Non-Local Thermodynamic Equilibrium (NLTE) effects into account, while at the same time radiative transfer effects often cannot be ignored. To account for the asymmetric morphologies of SNe, 3D input ejecta models must be used. In this work, we present the \texttt{ExTraSS} (EXplosive TRAnsient Spectral Simulator) code, which has been upgraded to be fully capable of 3D NLTE radiative transfer calculations in order to generate synthetic spectra for explosive transients in the nebular phase, with a focus on supernovae. We solve a long-standing difficulty of 3D NLTE radiative transfer -- to manage generation and storage of millions of photoexcitation rates over $\gtrsim10^{5}$ of cells -- by developing a new Domain Decomposition algorithm. We describe this new methodology and general code operations in detail, and analyse convergence and accuracy for \texttt{ExTraSS}. 

\end{abstract}




\section{Introduction}
The main goal of modelling supernova (SN) spectra is to determine key characteristics of the explosion, such as the ejected mass, the composition, and the explosion energy. These can be used to connect back to progenitor systems and obtain information on the explosion mechanism. These modelling efforts come in various flavours of complexity and with different focus points \citep[see][for a recent review]{jerkstrand2025spectral}. 

One example is codes assuming Local Thermodynamic Equilibrium (LTE) and modelling the outer line forming domain at early phases (e.g. \texttt{SYNAPPS}, \citealt{branch1980synthetic,thomas2011synapps}, the \citealt{mazzali1993application} code, and \texttt{TARDIS}, \citealt{kerzendorf2014spectral}). Such codes have the advantage of fast run times and can explore large parameter spaces. More detailed modelling considering certain NLTE effects was also done early in 1D \citep[e.g.][]{baron1995nonLTE,hoeflich1996explosion}. In later phases when LTE breaks down, consideration of a large variety of physical processes, and the full domain, are needed. Much of the initial groundwork to these more complex codes was begun by \citet{axelrod1980late}, who outlined the non-thermal and Non-Local Thermodynamic Equilibrium (NLTE) physics needed. Initially, the more complex codes were limited to either improved microphysics, or accounting for the multidimensionality of SNe. Nebular-phase (NLTE) codes in 1D were developed by \citet{fransson1989late,kozma1998late,mazzali2001nebular}, which operated without radiative transfer (in the optically thin limit). Early 2D work of the same type was done by \citet{mazzali2005asymmetric,maeda2006nebular}, in the context of modelling SN~2003jd and SN~1998bw. The first 3D LTE codes \texttt{SEDONA} \citep{kasen2006time} and \texttt{ARTIS} \citep{kromer2009time} slightly predate the second generation of 1D NLTE codes -- with radiative transfer -- such as \texttt{SUMO} \citep{jerkstrand201144ti,jerkstrand2012progenitor} and \texttt{CMFGEN} \citep[which was adapted to SN modelling by][]{hillier2012time}. Within the more complex codes, large variations still exist on how the spectral synthesis is performed, and which epoch of the SN is studied. Recently, many of these codes were compared by \citet{blondin2022standart}, which was a collaboration work by the different groups to investigate the level of agreement between their codes for a given input model. 

In the days and weeks after the explosion, the SN ejecta are in the so-called ``photospheric phase'', where the spectra probe the outer surface layers as optical depths are still high and radiation from deep layers cannot escape. In this phase, many absorption lines are observed, making detailed radiative transfer calculations critical, while the microphysics can be approached in a more simple manner (i.e. in LTE). At late times, the ejecta enter the ``nebular phase'', when the ejecta have become (mostly) optically thin and the inner regions are revealed. In this regime, optical depths are smaller, and dominant lines appear in emission rather than absorption. In this regime the radiative transfer becomes simpler to perform, but instead more microphysics (i.e. NLTE) must be accounted for. The transition into the nebular phase has no formally defined time and depends on the expansion velocity and mass of the ejecta, as well as the wavelength as redder emission becomes optically thin more quickly. A more detailed discussion can be found in \citet{jerkstrand2025spectral}.

SNe occur from two distinct progenitor systems: the thermonuclear explosion of (at least one) white dwarf (called Type Ia, SN Ia, see e.g. \citealt{taubenberger2017extremes} for an overview of the many different subtypes), and the core collapse of a massive star (CCSN; stars with $M_\text{ZAMS}\gtrsim8\,M_\odot$, \citealt{heger2003massive}, \citealt{jerkstrand2026core}) which encompasses all the other types, depending on what kind of star exploded \citep[see also][]{filippenko1997optical,galyam2017observational}. Multi-dimensional explosion simulations of SNe Ia have been available for quite some time \citep[see][for a review]{hillebrandt2013towards}, and a vast literature on 3D LTE modelling of these exists. NLTE modelling was done in 2D by \citet{botyanszki2018multidimensional}. \citet{shingles2020monte} started working towards including NLTE effects in the 3D code \texttt{ARTIS}, but computed models only in 1D. Recent studies which did progress into 3D focussed more on spectra up to $\sim2\,$months post-explosion \citep[e.g.][]{callan2025exploring,holas2025asymmetry,chen2025SEDONA} which -- even for SNe Ia -- is bordering on nebular phase, and done without detailed NLTE calculations. \citet{pollin2026multidimensional} did explore much deeper into the nebular phase, and implemented a modified photoionization treatment to compute NLTE effects.

In recent years, long-time hydrodynamical modelling of CCSNe has gained pace \citep[e.g.][]{wongwathanarat2015three,stockinger2020three,gabler2021infancy,vartanyan2025simulation,vartanyan2025simulated}. \citet[][\Jerk20 hereafter]{jerkstrand2020properties} performed the first processing of such a 3D explosion model, computing the deposition and emergence of gamma-ray lines. 
Spectral synthesis modelling in 3D NLTE began with \citet[][hereafter \vBPI23 and \vBPII24, respectively]{vanbaal2023modelling,vanbaal2024diagnostics}. These works used stripped star input models\footnote{Massive stars can lose their hydrogen envelope e.g. due to interactions with a binary companion, creating `stripped stars'. In more extreme cases, even the helium-rich envelope can be lost.}, which have lower ejecta masses and higher expansion velocities than hydrogen-rich CCSNe, and as such reach homologous expansion faster (but not as fast as SNe Ia). Due to this, their densities in the nebular phase are quite low and radiative transfer impacts are limited. \vBPI23 worked in a globally optically thin approximation, while \vBPII24 added an ``on-the-spot'' treatment for photoionization. For H-rich SNe (as modelled by e.g. \citealt{stockinger2020three}, \citealt{gabler2021infancy} and \citealt{vartanyan2025simulation}), radiative transfer effects are more important and must be included \citep{jerkstrand2012progenitor} -- the motivation for the current development work and application to Type II SNe \citep[see also][hereafter \vBPIII26]{vanbaal2026emission}. 

Including a complete radiative transfer method is computationally much more demanding than the optically thin version of \vBPI23. The photoexcitation rates alone might require upwards of 1~TB of data to store for a 3D grid with $\sim$100\,000 cells, depending on how many elements and excitation levels per ionization stage are used. Even on modern supercomputers, this is extremely challenging to achieve on a single node, and often unfeasible. As such, for 3D NLTE radiative transfer \emph{domain decomposition} must be pursued, to divide out the computational domain over the compute nodes. While having been developed and used for fission reactor simulations \citep[see e.g.][]{alme2001domain,brunner2009efficient}, and having been used in other astrophysical context several times \citep[see e.g.][]{hayek2010radiative,harries2019TORUS,micheldansac2020RASCAS}, the code version we present in this paper represents, to our knowledge, the first time this has been implemented for a supernova radiative transfer code. 

Domain decomposition generally is used for Monte Carlo transport problems which become too large to fit the data for the entire computational grid in the memory of a single core or node \citep{alme2001domain}. Instead, it becomes necessary to break the total grid into smaller subdomains, which reduces how much memory is required to be stored on each node, but which does come with additional communication overhead. The exact form of separation into the subdomains can have strong impacts on the efficiency of the scheme \citep{alme2001domain,brunner2009efficient}. An important question is to what degree transfer occurs between the different subdomains, as high `leakage' rates can lead to large communication costs and non-locality in the computational time \citep{siegel2012analysis}. 

In this work, we give a complete introduction and description to \texttt{ExTraSS} (EXplosive TRAnsient Spectral Simulator), earlier versions of which were used by \vBPI23 and \vBPII24, with the initial grid structure and $\gamma$-ray transport introduced by \Jerk20. The code has here been expanded to include global radiative transfer coupled to photoionization and photoexcitation, which is explained in detail, alongside the domain decomposition implementation. In a companion paper (\vBPIII26), the upgraded code is applied to the H-rich $9.0\,M_\odot$ model from \citet{stockinger2020three}. 

The paper is structured as follows: in Section~\ref{sec:methods}, we outline the different components of \texttt{ExTraSS} one by one, starting with the $\gamma$-ray transport and finishing with the radiative transfer and domain decomposition. In Section~\ref{sec:performance} we validate the stability and robustness of \texttt{ExTraSS} in its full form, \txtred{and in Section~\ref{sec:validation} we focus on the output,} specifically comparing how \texttt{ExTraSS} holds up compared to previous work and a benchmark testcase. In Section~\ref{sec:discussion} we explore the robustness of \texttt{ExTraSS}, some alternative approaches to the radiative transfer communication, and discuss future code updates. Our conclusion is given in Section~\ref{sec:conclusion}.

\section{Methods} \label{sec:methods}
\texttt{ExTraSS} uses a spherical coordinate grid. This has the advantage that cells at larger radii can be made larger while cells at inner radii can be small, enabling us to capture the fine structure at high resolution for the slower-moving ejecta, while retaining efficient transfer at large radii. The robustness of this spherical grid setup was shown in detail in \Jerk20 (their Appendix A). 

\texttt{ExTraSS} operates in snap-shot mode, meaning photon transport is followed through independent of time. Flight time effects can still be fully included for gamma-ray lines, as well as UVOIR lines if the non-local radiation field is ignored, following the methods of \Jerk20. For UVOIR lines with global radiative transfer those methods are insufficient, however. This means the code for this application is limited to relatively late phases when optical depths are low enough that the radiation transport time is unimportant. 

We discretize the angular variation of the radiation field with $N_{\theta\text{, VA}}\times N_{\phi\text{, VA}}$ viewers in the polar$\,\times\,$azimuthal directions. In the standard setup \texttt{ExTraSS} uses $N_{\theta\text{, VA}}=N_{\phi\text{, VA}}=20$, equally spread along both angles. As \texttt{ExTraSS} is designed with a spherical coordinate grid setup, this does mean that the viewers close to the poles (the smallest and largest $\theta$ angles) are `packed' more closely together, each representing a smaller part of the total sky angle. In the optically thin setups, this can be adjusted for as a post-processing step, while in ray-tracing calculations this has to be accounted for at emission, in order to ensure that the emission is sent off isotropically.

\begin{figure*}
    \centering
    \includegraphics[width=\linewidth]{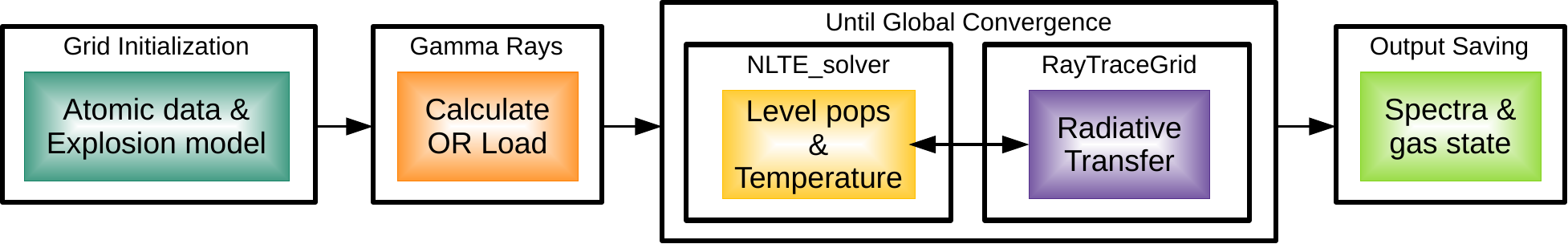}
    \vspace*{-.25cm}
    \caption{A flowchart to show the program flow of \texttt{ExTraSS} from the initial steps of model loading, to the final stages where the spectra and level population data are saved. Each box is compromised of several smaller subroutines and details, with the \textsc{NLTE\_solver} shown in Figure~\ref{fig:NLTE_schematic}.}
    \label{fig:TotalExTraSS_flowchart}
\end{figure*}
In this Section, we will first recap the technical details behind the $\gamma$-ray transport (\Jerk20), the NLTE solver (\vBPI23), and the radiative transport (the full version used with \vBPIII26; the optically thin version from \vBPI23 is described in Appendix~\ref{ssec:thin_spectra}, and the `on-the-spot' photoionization treatment from \vBPII24 in Appendix~\ref{ssec:OTSPI}). After this we will discuss the implementation of several memory reduction mechanisms including shared memory (introduced to the \texttt{ExTraSS} framework by \vBPI23) and then domain decomposition (introduced alongside the full Radiative Transfer treatment described here, and applied by \vBPIII26) -- in this last subsection we also discuss our implementation of domain decomposition in the context of previous astrophysical implementations. In Figure~\ref{fig:TotalExTraSS_flowchart}, a schematic overview is given for the structure of \texttt{ExTraSS}, with each box discussed in more detail in the subsections below.

The start of each run with \texttt{ExTraSS} is to initialize the atomic data and load the explosion model into the `grid' that is to be used. Additionally, a check is made to see if the $\gamma$-ray deposition can be loaded from a previous run, or if it has to be calculated first.

\subsection{$\gamma$-ray transport} \label{ssec:gammarays_method}
The $\gamma$-ray transport was first introduced in \Jerk20, who described several versions of their 3D photon packet transfer in order to study the radioactive decay lines of the $\gamma$-ray field. For the version of \texttt{ExTraSS} presented in this work, these original code setups can still be used, but they have also been integrated into the full 3D NLTE radiative transfer framework in various ways.

The $\gamma$-ray transport is by default computed using the `Compton scattering mode', from \Jerk20 (Sec~4.6). In this setting, a total number of packets $N_\gamma$ is divided across the cells which contain radioactive material ($^{56}$Ni,$^{56}$Co), scaled to the fraction f$_\text{cell}$ of the total $^{56}$Ni that is present in the whole model. Each core then sends a number of $\gamma$-ray packets from that cell $n_{\text{cell},\gamma}=N$$_\gamma\,\times\;$$\text{f}_\text{cell}(^{56}\text{Ni})/n_\text{core}$.

Each packet is given a random starting direction by isotropic sampling. For the transfer through the cells, the standard Monte Carlo formalism with co-moving frame energy transformations is applied \citep[e.g.][]{lucy2005monte,jerkstrand201144ti}, until scattering occurs. The photoelectric absorption effect is not included as it has negligible impact for energies $>68\,$keV \citep{alp2018xray}, and once a packet drops below $50\,$keV it is fully absorbed in the current cell. If a scattering occurs, part of the energy of the packet is lost to the cell where the scattering takes place, following the Compton formula. Upon scattering, the packets obtain a new travel direction. The reader is referred to \Jerk20 for more complete details. 

For every packet, random numbers are used to choose which of the 47 decay lines the $^{56}$Co will take, consistent with the branching ratios of these lines. Positrons (which carry $3.5\,\%$ of the energy on average) are assumed to be locally trapped. The $\gamma$-ray radiation field $J_\gamma$ is constructed following the method of \citet{lucy2005monte}, by following the path the packets traverse through the nebula.

\subsection{NLTE solver}  \label{ssec:NLTEsolver_method}
\begin{figure}
    \centering
    \includegraphics[width=\linewidth]{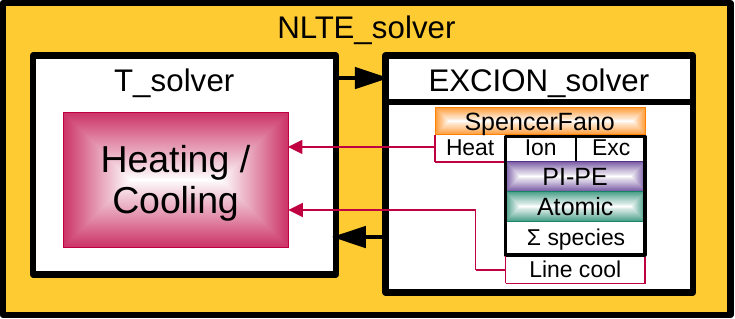}
    \vspace*{-.05cm}
    \caption{A flowchart to show the key pieces inside the \textsc{NLTE\_solver} subroutine. The \textsc{SpencerFano} subroutine makes use of the $\gamma$-ray energy deposition, the \textsc{PI-PE} block determines photoionization and photoexcitation from the radiative transfer, an the \textsc{Atomic} block generates the rates between all the level populations accounting for all atomic processes plus SF and RT effects. Inside the \textsc{T\_solver}, the heating from $\gamma$-rays and photoionization is balanced against bound-bound and free-bound cooling from the \textsc{EXCION\_solver}.}
\label{fig:NLTE_schematic}
\end{figure}

The NLTE level population solver was first introduced in \vBPI23 but has been updated and improved since. This \textsc{NLTE\_solver} will solve for the local temperature $T_\textbf{cell}$ and the NLTE level populations in an iterative manner, holding photoexcitation and photoionization rates fixed. Both thermal and non-thermal physics are accounted for in the solver, alongside impacts from the non-local radiation field. The main output of the \textsc{NLTE\_solver} are the level population solutions (see Figure~\ref{fig:TotalExTraSS_flowchart}), which generate the radiation field. Figure~\ref{fig:NLTE_schematic} shows a more detailed look at the most critical pieces of the solver and how they affect each other.

The total number of cells in an \texttt{ExTraSS} model is typically $\mathcal{O}(10^{5-6})$, which means that some care has to be taken in the design of the level population solver to keep this computationally feasible (both regarding runtime and RAM -- see also Sec~\ref{ssec:jaggedmemory_method}). As noted in \vBPI23, for every ion we include up to 100 excitation levels, with three ionization stages per element (except H), meaning up to $100\times N_\mathrm{ion stages}$ excitation stages are calculated per atomic species. Every atomic species is calculated separately, but the excitation and ionization structures are calculated concurrently in the so-called \textsc{EXCION\_solver} (the right box in Figure~\ref{fig:NLTE_schematic}). \texttt{ExTraSS} has atomic data sourced from \texttt{SUMO}, currently including H, He, C, N, O, Ne, Na, Mg, Al, Si, S, Ar, K, Ca, Sc, Ti, V, Cr, Mn, Fe, Co and Ni\footnote{These are the elements present in the typical \texttt{P-HotB} long-time simulation models that have been employed as inputs, alongside the important odd-Z elements. Some of those elements, such as Na and N, have lines in the optical in the nebular phase \citep[see e.g.][]{jerkstrand2018emission,dessart2021nebular,barmentloo2024nebular} although they are not usually included in the input models employed by \texttt{ExTraSS} so far.}, meaning up to 22 unique atomic species might be solved for per cell. When atomic data for collision strengths, photoionization cross sections, or recombination rates are missing, the same approximate treatments are done as in \texttt{SUMO}. 

As part of the \textsc{NLTE\_solver}, we employ the Spencer-Fano subroutine of \citet{kozma1992gamma} to determine the distribution of the non-thermal electrons from the $\gamma$-ray energy deposition (orange boxes in Figures~\ref{fig:TotalExTraSS_flowchart} and \ref{fig:NLTE_schematic}). In the current version of \texttt{ExTraSS}, all three channels of heating, ionization and excitation are included. As the inclusion of the excitation channel makes the Spencer-Fano calculation computationally expensive, it is only called on the first and fourth (\textsc{NLTE\_solver}) iteration, and every subsequent ten iterations afterwards. Additionally, if all elements except H have reached convergence, it is no longer called\footnote{Tests indicated that occasionally H will take many more iterations in the \textsc{EXCION\_solver} than other elements, and that subsequent Spencer-Fano calls had no impact on the final solution but did significantly impact the runtime.}. In \vBPI23 and \vBPII24, the excitation channel was not used, and instead the Spencer-Fano routine was called the first two iterations and then every 4th.

The \textsc{NLTE\_solver} also accounts for the impact of the radiation field through photoionization and photoexcitation (the purple `PI-PE' box in Figure~\ref{fig:NLTE_schematic}, see also Section~\ref{ssec:RadTrans_method}). The first global iteration, the radiation field has not been calculated yet, and instead then the \textsc{NLTE\_solver} applies the local `on-the-spot' photoionization treatment from \vBPII24 (see Appendix~\ref{ssec:OTSPI} for a description of this treatment). This treatment takes recombination emission and applies the hydrogenic approximation to determine photoionization cross sections for all ground state multiplets plus the next eight excited states. The excess energy from the ionization is accounted for as extra heating source.

In total, the \textsc{NLTE\_solver} will do up to 150 iterations which each exist of one call to the \textsc{T\_solver} and one call to the \textsc{EXCION\_solver}; afterwards the temperature is adjusted, accounting for collisional bound-bound emission, free-bound emission, heating through photoionization and non-thermal electron heating (through Spencer-Fano). The maximum d$T_\text{max}$ per \textsc{T\_solver} is $\text{damp}\times250\,$K, with the damping factor initially at 0.8, decreasing to 0.2 after 75 iterations and 0.02 after 110. The temperature update is constructed such that back-and-forth jumping is avoided\footnote{In trials, it was noted that sometimes $T_\text{cell}$ could jump between two values (e.g. 3100\,K and 2900\,K) until the damping factor reduced d$T_\text{max}$. If such a looping would occur, the new $T_\text{cell}$ is set to 75\% of the higher value plus 25\% of the lower one.}. Convergence is achieved when d$T_\text{prev}$ is less than 0.1\% of the current $T_\text{cell}$ and if the `net heating rate' is less than 5\% of the total heating rate.

The \textsc{EXCION\_solver} determines the level populations of all present elements consecutively. This is done through a matrix calculation, containing all the rates between all the levels, and constructing a numerical Jacobian to update to the new level populations. These are calculated through a Newton-Raphson scheme, with a decreasing damping scheme starting at damp $=0.95$ and decreasing to 0.75, 0.5, 0.25 and 0.05, after 10, 35, 50 and 90 iterations. An atomic species is considered converged if the difference between the old and new level populations is less than $1\%$ for every state.

In \vBPI23 and \vBPII24, there were no non-local effects to consider for the cells, and as such once the \textsc{NLTE\_solver} converged to a solution, the cell as a whole can be considered to have converged. Convergence failure is quite rare, but can occur, typically in cells with lower densities and with low energy deposition. Cells that fail to converge are excluded in the spectral synthesis. A detailed description for the optically thin emission is given in Appendix~\ref{ssec:thin_spectra}.

\subsection{Radiative transfer} \label{ssec:RadTrans_method}
In order to capture effects from the globally sourced radiation field, full radiative transport has to be implemented. To achieve this, the `on-the-spot' treatment from \vBPII24 is expanded, to include the full wavelength range and to use more levels from which photoionization (see Section~\ref{sssec:PI_RT}) occurs. Additionally, photoexcitation (Section~\ref{sssec:PE_RT}) rates have been added to account for lines.

The radiation field is both generated by the level populations in each cell, and interacts with them through photoionization and photoexcitation, so this mandates that \texttt{ExTraSS} iterates between the \textsc{NLTE\_solver} and the radiative transport module (\textsc{RayTraceGrid}), as shown in Figure~\ref{fig:TotalExTraSS_flowchart}. 

On each node, one core is assigned as manager, and this core will handle the data transfer between different nodes (see also Section~\ref{sssec:dodec_method}). All other cores, or `workers', are responsible for generating and transporting the rays from the level population solutions from the \textsc{NLTE\_solver}. For this, the mechanisms of bound-bound emission, free-bound emission and two-photon emission for neutral H \citep{nussbaumer1984hydrogenic} and He \citep{li1995hetwophoton} are considered. Subsequently, the emission is binned by wavelength ($400-25000\,\angstrom$, with logarithmic step sizes of $0.1\,\%$) to generate rays. Emission blueward of $15000\,\angstrom$ is treated with the new radiative transport, while emission between $15000-25000\,\angstrom$ is treated as before under the optically thin treatment -- the transition point for the optically thin emission is flexible. Emission outside of the wavelength range is tracked for convergence checks. 

We use a ray tracing technique to follow the path of the photon packets from their emission point, which is the middle of the cell, until they either escape the grid or less than $10^{-6}$ of their starting energy remains. Rays are emitted towards all viewing angles; the default setup uses $20\times20$ viewers (as in \Jerk20,\vBPI23,\vBPII24,\vBPIII26) equally spread in the polar ($\theta$) and azimuthal ($\phi$) angles. The 400 beam directions give a satisfactory angular sampling to obtain good photoexcitation and photoionization rates in each cell.

The ray tracing is done in a similar manner as described in (\Jerk20, their Sec~4.5), but with modifications to the optical depths and the division of emission into the energy packets ($j_i$ in \Jerk20). For the optical depths, these are now calculated with respect to photoionization (see below, also for photoexcitation calculation details) and not for Compton scattering. Meanwhile, the energy packet subdivision is altered to adjust for the solid angle of the viewing direction ($\Delta\Omega_k$ in \Jerk20). This is done because the viewing directions are not spread isotropically\footnote{With a setup of $20\times20$ viewing angles for the polar$\,\times\,$equatorial angles, viewing angles closer to the north/south poles are packed more tightly and cover a smaller patch of the sky.}, which means without such a correction the emission would not be isotropic as it should be. The solid angle is also accounted for when generating the final output spectra (\Jerk20 did no deposition calculations in the ray-tracing mode, so the solid angles were pre-cancelled).

In each cell $i$, for each wavelength bin $j$, one ray will be sent towards each observer $k$; thus the total number of rays can be computed ahead of time and used to verify when the radiative transfer has completed. Each ray has a number of photons $N^\text{phot}_{i,j,k} = E^{\text{bin}}_{i,j} / \Delta\Omega_k / E^\text{phot}_{j}$, where $E^\text{phot}_j$ is the energy ($hc/\lambda_j$) of an individual photon at the corresponding wavelength $\lambda_j$ where the ray starts and $E^\text{bin}_{i,j}$ is the starting emission in this bin and cell. The starting $\lambda_j$ is chosen to correspond to the reddest $\lambda$ of all features within the bin $j$ (to avoid self-absorptions). For the strong lines Mg~I] $\lambda\,4571$, [O~I] $\lambda\lambda\,6300,\,6364$, H$\alpha$ (including $nl$ splitting), [Ca~II] $\lambda\lambda\,7291,\,7323$, Ca~II $\lambda\lambda\lambda\,8498,\,8542,\,8662$ and [C~I] $\lambda\,8727$, a narrow emission bracket is created to make sure these features are emitted at the right wavelength and not blended with any other emissions. Although the brackets we create for these emission features are actually smaller than the thermal widths of their lines, any thermal motion from within one cell will be lost among the Doppler broadening that will take place from all cells together.

Once the ray tracing has finished, the rates for photoionization and photoexcitation can be finalized (details below), and these rates are coupled back to the \textsc{NLTE\_solver} to be included in the next iteration. Global convergence is checked by comparing the total energy in the emergent spectra for all viewing angles (as well as any escaped emission further redwards of $25000\,\angstrom$) to the total energy deposition inside the nebula, as the ratio between these should be close to unity when in the steady-state mode (the transfer introduces some adiabatic losses in a homologous flow, so the typical value is in the range of 0.9--1). Electron scattering is currently not treated in the code, which means we do not capture the line peak blueshifting and red tail enhancement that can occur in the early nebular phase.

\subsubsection{Photoionization} \label{sssec:PI_RT}
Every time a ray path segment is traced -- which is either to the wall of the next cell, or to a line (in the Sobolev approximation, line interactions happen at discrete resonance points) -- \texttt{ExTraSS} computes the optical depths $\tau_\text{PI}$ for photoionization. For calculations involving the ground state, data from \citet{verner1996atomic} is used, while for calculations involving excited states the hydrogenic approximation is used (see also Equation~\ref{eq:sigma_OTSPI}). \txtred{For a few excited levels, cross sections from TOPBASE are used (H~I $n=2$, He~I $2^{1}$S and $2^1$P$^0$, N~I $^2$D, O~I $^1$D and $^1$S, Mg~I $^3$P, Mg~II $^2$P, S~I $^1$D, Ca~II $^2$D, Fe~I a$^5$F and Fe~II a$^4$F -- see also \citealt{jerkstrand201144ti}).}

The number of levels which are included in the photoionization calculations varies per ion. This is done as the calculations of $\sigma_\text{PI}$ are expensive, yet for some species it is critical that enough levels are included (e.g. O~I, Ca~I, Fe~I) while for others only one or a few levels have to be accounted for (e.g. He~I, Ar~I). Currently, \texttt{ExTraSS} accounts for neutral, singly ionized and doubly ionized -- so there is no ionization calculated from doubly ionized upwards. The number of levels are chosen such that enough levels are included to obtain accurate photoionization rates, while limiting the number of $\sigma_\text{PI}$ values that have to be calculated, for computational reasons.

For each distance $d$ that a ray travels, $\sigma_\text{PI}$ is calculated for all levels which have a lower ionization energy threshold than the energy of the photons in the ray. \txtred{If the ray shifts past the wavelength threshold, $d$ is modified to account for this.} These are then combined to get an optical depth $\tau_{\text{PI},i} = \sigma_{\text{PI},i}\ \times\ d\ \times\ n_i$ for all levels $i$, where $n_i$ are the level populations of all considered levels for all ions. These $\tau_{\text{PI},i}$ are then summed together to determine the total number of photons absorbed $N_\text{abs} = N_\text{phot}\times(1-\mathrm{e}^{-\tau})$, with each level $i$ absorbing $N_\text{abs}\times\tau_{\text{PI},i} / \Sigma_i\tau_{\text{PI},i}$ photons. The excess energy between the photon energy and the ionization energy is accounted for as heating source.

At the end of the \textsc{RayTraceGrid} calculation, the number of absorbed photons for each level $i$ are converted to a photoionization rate $\Gamma_{\text{PI},i}=N_{\text{abs},i} / N_i$, where $N_{\text{abs},i}$ is the number of photons absorbed by level $i$ and $N_i$ is the total number of particles in that level (so the level population $n_i$ times the volume of the cell). These rates are calculated separately for all cells. We do not consider effects of stimulated recombination.

\subsubsection{Photoexcitation} \label{sssec:PE_RT}
Every step a ray makes through the grid which covers some distance $d$, the wavelength of the ray is updated to the co-moving frame. To check for line interaction, a check is made to see if $\lambda_\text{start} < \lambda_\text{line} < \lambda_\text{end}$ for the individual line with the closest wavelength to $\lambda_\text{start}$. All lines which have an optical depth of $\tau_\text{line}>10^{-4}$ in the cell that is being traversed are considered possible interaction points.

If line interaction takes place, then the ray will move the distance $d_\text{line}$ instead, stopping inside the cell, and perform the photoionization calculation for this shorter distance $d_\text{line}$. If the ray has not been fully lost to photoionization, the number of photons absorbed by the line is calculated through $N_\text{PE}=N_\text{phot}\times (1-\mathrm{e}^{-\tau_\text{line}})/f_\mathrm{corr}$, where $N_\text{phot}$ is the total number of photons in the ray and $f_\mathrm{corr}$ is a correction for stimulated emission\footnote{See Appendix B.1 in \citet{jerkstrand2012progenitor}.}. \txtred{After the line interaction, the ray will continue on its path, provided that the ray is still above the $10^{-6}$ starting energy threshold, which is checked by $(N_\text{phot}-N_\text{PE})\times E_\text{line}$, where $E_\text{line}$ is the energy of a photon at the wavelength of the line.} At the end of the \textsc{RayTraceGrid} calculation, the photons absorbed in each line in each cell are converted to a photoexcitation rate, and a photodeexcitation rate is determined from detailed balance.

\subsection{Memory optimization} \label{ssec:jaggedmemory_method}
Just storing the level populations across $\sim10^6\,$cells, for 65 ions with a maximum of 100 excitation levels ($N_\text{lev}$) each would already require $\sim6.5\times10^9$ numbers, which is $\sim50\,$GB of RAM. Storing the photoexcitation rates would take another factor of $N_\text{lev}$; this would clearly result in a problematic amount of data to hold in memory. \texttt{ExTraSS} was therefore designed with a strong focus on optimizing memory usage.

First, we rely heavily on shared memory between cores on one node. This includes -- but is not limited to -- large parts of the atomic dataset, critical cell data, and spectral output. Second, \texttt{ExTraSS} makes use of ``Jagged Arrays'', in order to only allocate memory when it is needed. For example, not all ion species have (at least) 100 excitation stages, so for many less data can be used. In particular for photoexcitation calculations, which scale with $N_\text{lev}^2$, this reduces the memory load significantly. Third, only if an element is present in a cell, data gets allocated to store its level population, further reducing memory usage.

However, even with the combination of Jagged Arrays and shared memory, full 3D NLTE Radiative Transfer on large models will require a large amount of RAM. Further steps towards tractability requires a domain decomposition scheme which cuts the full problem into several smaller ones, and each node computing only its assigned part. By only loading this part, the total memory usage on each node goes down, and will scale proportional to how many nodes are being used -- if the problem is too big, introducing more nodes will now reduce the memory requirements. Domain decomposition removes the roof on how detailed microphysics can be treated for a given problem set by node memory limits, opening a path towards highly accurate modelling. 

\subsubsection{Domain decomposition} \label{sssec:dodec_method}
\begin{figure}
    \centering
    \includegraphics[width=\linewidth]{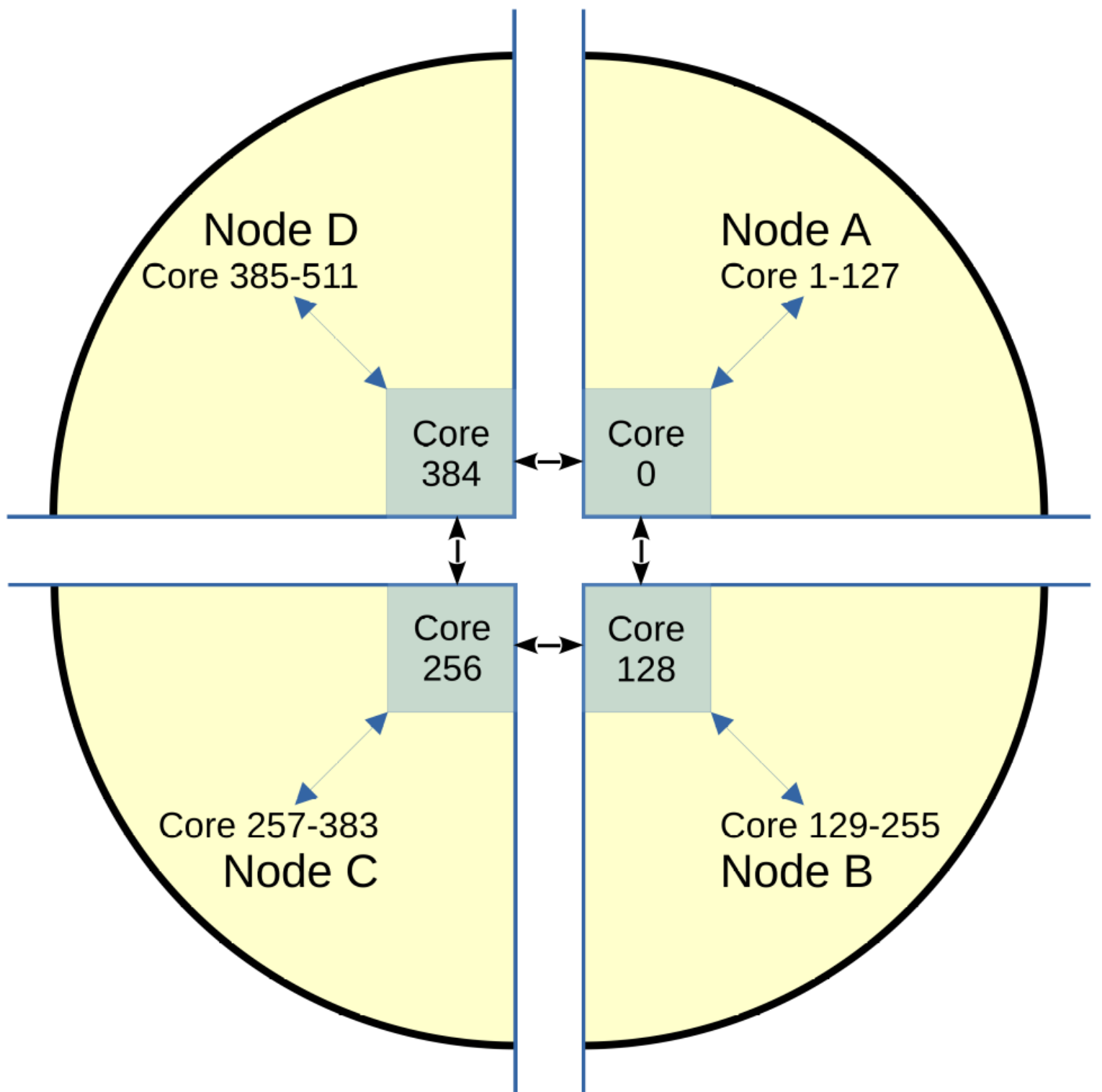}
    \caption{An equatorial slice of the 3D grid, to show how this is broken into four slices by the domain decomposition scheme, with the main communicating cores on each node marked, and the communication chains represented by the arrows. In this example, the nodes each have 128 cores, but this number will vary for different clusters.}
    \label{fig:DoDec_sketch}
\end{figure}
The domain decomposition scheme implemented in \texttt{ExTraSS} is inspired by \citet{brunner2009efficient}, who outlined an algorithm which makes use of non-blocking communication for both sending and receiving data. In our setup, the communication between domains is managed by one core on each node, which is sufficient to not overload this core which would slow down the whole communication stream. The goal of the domain decomposition technique is to reduce memory load on the system at the cost of some communication overhead. This communication overhead can be reduced by properly splitting the full domain into smaller ones, which in our case means $\phi$-based decomposition, i.e. each node takes an equal part of the number of azimuthal slices to make a domain. 

The first component in a domain decomposition scheme is to set up the data communication between the different domains. In \texttt{ExTraSS}, this is done by having one core (the `manager') on each node assigned to manage both the intra- and inter-node communication, while the rest of the cores generate and transport the rays. The manager cores talk both to each other (so cross-domain) to handle data transfer between the domains, as well as to all other cores on their node to assign incoming rays to new workers and to receive rays which need to be transferred to a different domain (bundles from emissivity are generated by workers themselves). 

Transfer of data between cores on the same node is highly efficient and quick, while cross-node transfer is slower to initiate and should be done in fewer but more data-intensive steps. Optimization of bundle sizes for both types of transfer was done (see below), but it should be noted that differently sized grids and number of nodes will have different ideal sizes. In order to retain good scalability for the problem, communication is set up asynchronously, as this optimizes worker efficiency \citep{brunner2009efficient}.

In Figure~\ref{fig:DoDec_sketch}, we schematically display the separation of a 2D circular grid (or, equivalently, an equatorial slice of a 3D spherical grid) into four equal domains. The arrows in Figure~\ref{fig:DoDec_sketch} outline the communication chains that are initialized alongside this domain-separation. The number of used domains within \texttt{ExTraSS} is flexible; the representation with four slices is visually the cleanest while still outlining that only communication between neighbouring nodes is initialized.

\begin{figure}
    \centering
    \includegraphics[width=\linewidth]{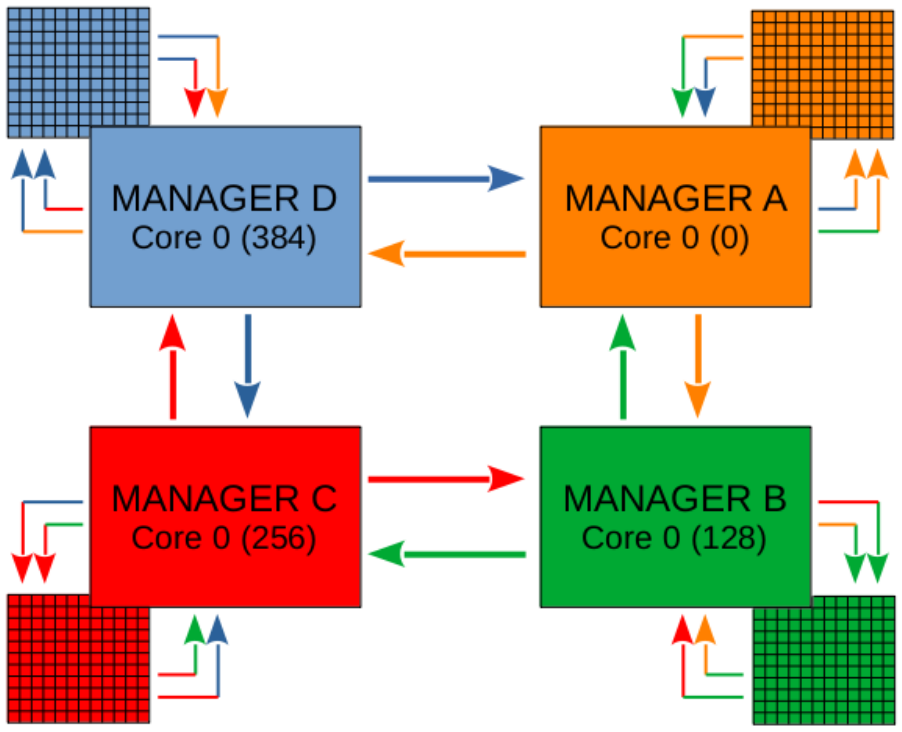}
    \caption{A schematic overview of all the communication streams for the four-domain setup from Figure~\ref{fig:DoDec_sketch}. The grid of small blocks represents the 127 workers (cores 1-127 on each node) per node, the small arrows communication between each worker and their manager, and the bigger arrows between the different managers. Each node has its own colour, with the small arrows indicating both the ``origin'' of the data (base) as well as the ``target'' (head). Each arrow represents a different MPI communication stream -- the small arrows represent all 127 streams between the workers and their manager.}
    \label{fig:MPIcomm}
\end{figure}

A more detailed look at the communication chains initialized for the domain decomposition is shown in Figure~\ref{fig:MPIcomm}. As shown by the amount of arrows, the data transfer between the managers is handled separately for the `up' and `down' streams. 

We verified the robustness of the code both for the total number of domains used and for the size of the communication streams from the managers (both to other managers and to their own workers). For the current work, the domain decomposition is limited to only separating by $\phi$-angle (as is shown in Figure~\ref{fig:DoDec_sketch}). The $\phi$-slicing is the simplest to implement, minimizes domain crossings, and is optimal for maintaining (roughly) equal domain loads. Intra-node communication is more efficient, and thus data bundles between workers and managers are smaller than between two managers, which is represented by arrow size in Figure~\ref{fig:MPIcomm}. 

The optimization of these bundle sizes depends quite strongly on the number of cores available on a node, the number of domains and to a lesser degree on the total number of rays that will be used in the radiative transfer. For manager-to-worker communication, a bundle size of 65536 ($=2^{16}$) was chosen, while manager-to-manager data bundles are 24 times larger, i.e. the manager cores can receive 24 data bundles from their workers before they initiate data transfer to the target node, and when they receive data from another manager they split this back down into 24 smaller bundles to hand back to their own workers. Some details on the testing of these bundles sizes is shown in Appendix~\ref{app:bundletesting}.

Other astrophysical codes which make use of domain decomposition typically use blocking MPI communication \citep[e.g.][]{harries2019TORUS,micheldansac2020RASCAS}. With \texttt{ExTraSS}, we opted for non-blocking communication, as we want to ensure that the `workers' spend as little time as possible waiting for communication chains, and instead spend their time computing ray tracing. The implementation by \citet{micheldansac2020RASCAS} included `ghost cells', i.e. boundary regions between two nodes that both nodes are aware of, in order to reduce data bouncing back and forth between two nodes. For the current version of \texttt{ExTraSS}, where rays do not undergo any scattering inside the \textsc{RayTraceGrid} module\footnote{For the $\gamma$-ray calculations, which is handled separately from this transfer, scattering \emph{is} included.}, there is no need to add such `ghost cells'. However, implementing such boundary regions should be considered if scattering is added in future versions.

\section{Code \txtred{Performance}}  \label{sec:performance}
In this Section, we discuss and validate several of our setups chosen for the different modules inside \texttt{ExTraSS}. Comparisons to other works and codes will be shown in Section~\ref{sec:validation}.

The validation tests were performed on the CPU partition of the Dardel HPE Cray EX system. The tests were done using nodes which have $512\,$GB of RAM and two AMD EPYC\texttrademark{} Zen2 $2.25\,$GHz 64-core processors, giving each node 128 physical CPU cores.

The model used for the validation runs is the 3D model from \vBPIII26 at 400 days, which is composed of $38\,\times\,15\,\times\,30$ zones (for $N_r,N_\theta,N_\phi$). \txtred{Despite its relatively low cell count, this model covers a good} part of the total possible parameter space in terms of composition, density and energy deposition. The model from \vBPIII26 contains fewer cells than the He-core models from \vBPII24, but does include H which is important for the testing.

\subsection{NLTE solver setup}  \label{ssec:NLTEsolver_valid}
Within the \textsc{NLTE\_solver} framework, there are two main computational factors. The first is the Spencer-Fano subroutine, to determine the degradation spectrum of the Compton electrons, and the second is the matrix solver used to calculate the new level populations. The Spencer-Fano subroutine includes non-thermal excitation, ionization and heating terms. Technically, every time the level populations have been adjusted within the \textsc{EXCION\_solver} routine, the degradation spectrum and associated rates should be re-calculated. In practice, this would be extremely expensive, and instead a scheme can be devised to limit the amount of Spencer-Fano calls while finding converged and accurate solutions.

\begin{figure}
    \centering
    \includegraphics[width=\linewidth]{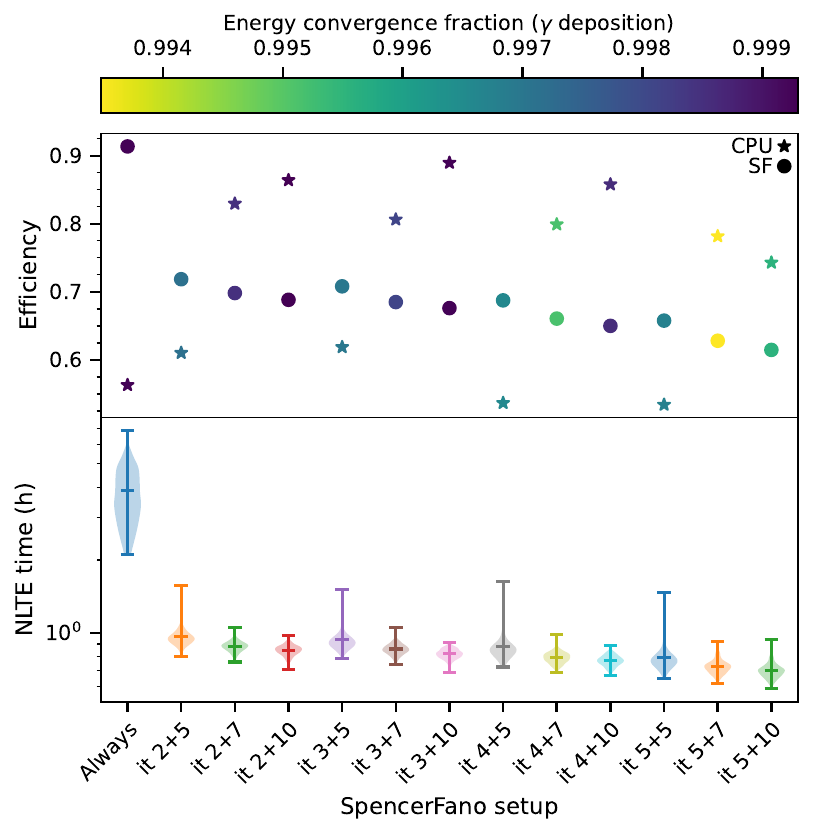}
    \caption{An overview of the computational and convergence efficiency of 13 different setups for how often Spencer-Fano subroutine is called (x-axis labels; the +X indicates the subroutine is repeated every X iterations). The top panel shows the efficiency in terms of CPU uptime (stars) and for the fraction of the total runtime spent inside the Spencer-Fano routine (circles), while the bottom panel shows violin plots of the time spend in the \textsc{NLTE\_solver} for each individual worker core. The markers in the top panel are colour coded by the fraction of $\gamma$-ray energy deposition in all converged cells, with darker colours having more total energy convergence.}
    \label{fig:SFNLTE_stab}
\end{figure}
\subsubsection{Spencer-Fano schemes}
Whichever scheme is devised to optimize the runtime and convergence stability, the \textsc{EXCION\_solver} will end up calling Spencer-Fano only every so often. In Figure~\ref{fig:SFNLTE_stab} the outcomes of 12 different schemes are shown alongside the setting where Spencer-Fano is called every iteration (labelled `Always'). In every other scheme (labelled `it A+X'), Spencer-Fano is called on iterations 1, A, and then every X'th iteration from A onwards (e.g. 2+5 calls at 1, 2, 7, 12, etc.). Every run for this comparison made use of 3 CPU nodes of the Dardel cluster (for 384 total cores). \txtred{Across all setups, the same cells are solved by the same cores.}

The top panel of Figure~\ref{fig:SFNLTE_stab} shows the `Efficiency' for the CPU-uptime (marked by the stars), and for the fraction of the total computational time spent on the Spencer-Fano subroutine (circles). Ideal load-balancing schemes would have a CPU-time of (close to) 1, so the good setups here tend to be the +7 and +10 schemes, while calling Spencer-Fano every iteration leads to one of the lowest load-balancing setups. Calling Spencer-Fano less regularly and starting later leads to less time spent inside this routine, but even in the scheme which makes the fewest calls (5+10), Spencer-Fano ends up taking more than $60\,\%$ of the total computational time of the \textsc{NLTE\_solver}. The colour coding in the top panel is set by taking the energy of all converged cells, compared to the total energy in the model. Even in the worst schemes, more than $99\,\%$ of the energy resides in converged cells, but ideal setups approach total energy convergence. Schemes that call Spencer-Fano every +5 and +7 tend to be less robust than the +10 setups, which approach similar convergence success as the `Always' setup.

The bottom panel of Figure~\ref{fig:SFNLTE_stab} clearly shows that calling Spencer-Fano every iteration is needlessly expensive, with mean runtimes per core about 4x as long as in all the other schemes, and the slowest core is about 3.5 times slower than the fastest core in this setup, indicating poor load balancing. Relatively poor load balancing can also be observed for the +5 schemes, with a ratio of about two between the fastest and slowest cores. The schemes with +7 and +10 show much better load balancing, with the +7 schemes having occasional slow outliers while the +10 schemes tend to have fast outliers. \txtred{The slow outliers in the +5 schemes are not consistently the same cores, as there are 18 unique cores among the slowest 5 per scheme. The slowest core (a different one for each setup) is some $10-15\,$minutes slower than the 5th slowest one ($\sim20\,\%$ of the NLTE runtime). That these outlier cores vary so much might indicate that this is not (purely) a model quirk, but rather intrinsic behaviour for the \textsc{EXCION\_solver} -- if many more iterations occur in the +5 schemes for particularly low energy deposition/low density cells, it would lead to patterns like these although even then it might have been expected that the same core is always the culprit, which is not the case.} 

Combining the runtime efficiency with the convergence stability, the default setting for \texttt{ExTraSS} is to use the 4+10 scheme, which has a CPU-time efficiency of $\sim0.85$, an energy convergence fraction of $0.9986$ and a mean \textsc{NLTE\_solver} runtime of 46 minutes (with each core calculating 45 cells) for the test model.

\subsubsection{\textsc{EXCION\_solver} matrix schemes}
The second main computational cost in the \textsc{NLTE\_solver} resides in the matrix calculation to determine the new level populations. For particularly small atoms (e.g. Ne, Ar can be computed while accounting for only a few excited states in each ion), this matrix calculation will have negligible cost, while for larger atoms (e.g. Ti, Fe, Co) this can become expensive. Several parameters can be tweaked within the \textsc{EXCION\_solver} to impact the solution time: the damping factor of the Newton-Raphson scheme, the maximum variation in level populations between two steps, or the maximum number of excitation levels used per ion.

\begin{figure}
    \centering
    \includegraphics[width=\linewidth]{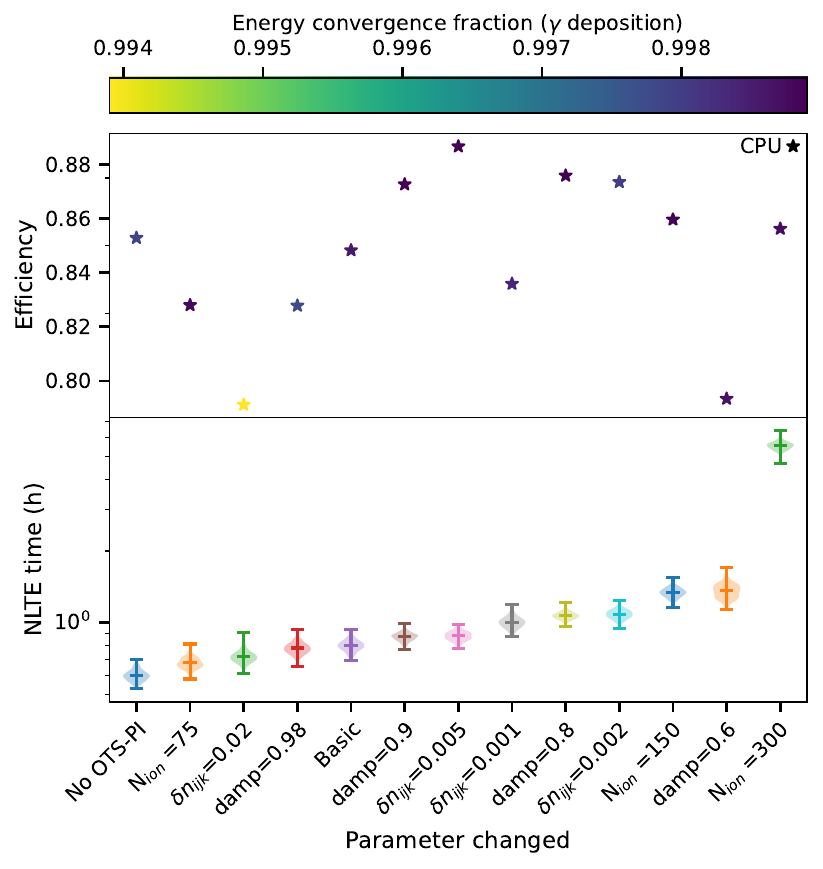}
    \caption{An overview of the computational and convergence efficiency of 13 variations of the \textsc{EXCION\_solver} setup (see x-axis labels), sorted by their total runtimes (lowest on the left, highest on the right). `OTS-PI' refers to the `on-the-spot' photoionization scheme (see Appendix~\ref{ssec:OTSPI}), $N_\text{ion} = X$ where $X$ is the maximum number of excited levels per ion, $\delta n_{ijk}$ indicates maximum difference between previous and new level populations to allow convergence, and damp\,$=X$ with $X$ as the damping factor in the Newton-Raphson scheme. The top panel shows the efficiency in terms of CPU uptime (stars), while the bottom panel shows violin plots of the time spend in the \textsc{NLTE\_solver} for each individual worker core. The markers in the top panel are colour coded by the fraction of $\gamma$-ray energy deposition in all converged cells, with darker colours having more total energy convergence.}
    \label{fig:EXCION_stab}
\end{figure}

In Figure~\ref{fig:EXCION_stab}, the robustness of these three options is compared to the basic setup, plus a setup where the `on-the-spot' photoionization mechanism (see Appendix~\ref{ssec:OTSPI}) was disabled. The setups are sorted from left to right by total runtime, with the top panel showing the CPU efficiency alongside colour coding for total energy convergence within the model. For the `basic' setup, $N_\text{ion}=100$, $\delta n_{ijk}=0.01$ and damp\,$=0.95$; all other setups varied one of these parameters. Each run made use of 3 CPU nodes of the Dardel cluster.

Compared to the CPU uptime for the various Spencer-Fano tests, the other variations within the \textsc{EXCION\_solver} lead to much smaller fluctuations for the CPU uptime, with the worst load-balancing setups still approaching $80\,\%$ efficiency. Additionally, only one setup has a noticeably low energy convergence ($\delta n_{ijk}=0.02$). Setup changes which tend to simplify the solver also make it faster -- lower number of levels per ion, more level population difference allowed between iterations, and a higher damping factor, although they also (somewhat) reduce the energy convergence. Additionally, lowering the number of levels per ion can also have severe consequences down the line, as it reduces the accuracy of the atomic data. The simplification of removing the `on-the-spot' photoionization has the largest immediate impact on the runtime, but this does also cause strong impacts on the level population solutions which then cascades into the radiative transfer, and could cause the need for more global iterations before global convergence is achieved. On the other side, more complex setups (significantly) slow down the solver, with $N_\text{ion}=300$ taking almost 7 times as long as the basic setup although some of these are also among the setups with the best energy convergence.

Within the model used here, several of these altered setup perform similarly, or perhaps even somewhat better, than the basic setup. However, what setup works best does vary per model, and for the He-core models from \vBPI23 and \vBPII24, altering the damping factor tended to decrease the total stability noticeably, leading to lower energy convergence rates, while for the model here only the $\delta n_{ijk}$ shows this behaviour. Generally, the similarity of the runtimes, CPU efficiency and energy convergence between many of these setups indicates the robustness of the \textsc{NLTE\_solver} as a whole.

\subsection{Radiative transfer setup}  \label{ssec:RadTrans_valid}
\txtred{Here, we focus on the computational performance of the radiative transfer setup, while the outputs and calculated rates are shown and investigated in Section~\ref{sec:validation}.}

\begin{figure}
    \centering
    \includegraphics[width=\linewidth]{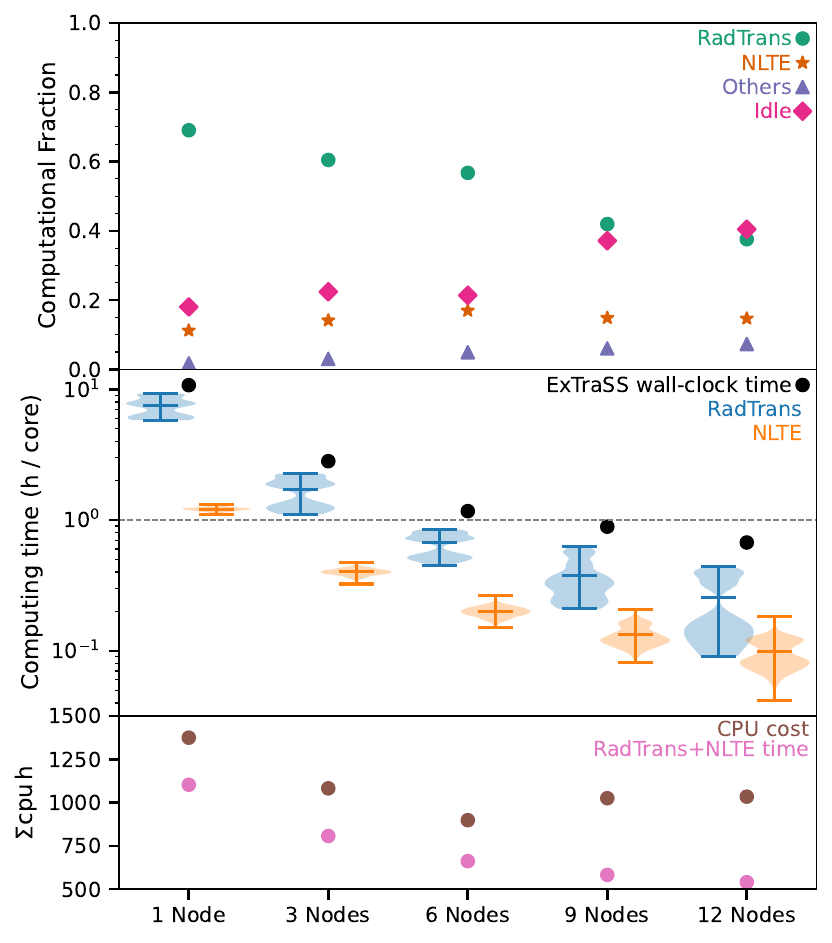}
    \caption{An overview of the computational efficiency for the radiative transfer (\textsc{RadTrans}) module, the \textsc{NLTE\_solver}, and the total \texttt{ExTraSS} runtime for an increasing number of nodes (see x-axis labels), when running the \textsc{RayTraceGrid} and \textsc{NLTE\_solver} modules once each, from the non-converged starting solution for the model from \vBPIII26. The full radiative transfer calculation computes $2.5\times10^{10}$ rays. The top panel shows the computational fraction (time spent relative to the total \texttt{ExTraSS} runtime) of the radiative transfer module (green circles), the NLTE solver module (orange stars), intermediate computational tasks (purple triangles), and the core-idling time (fuchsia diamonds). In the middle panel, we showcase the spread of the runtimes per core as violin plots, with bars indicating the min, mean and max values, for the \textsc{RayTraceGrid} module (blue) and the \textsc{NLTE\_solver} (orange), with the black dot indicating the wall-clock time for \texttt{ExTraSS}. The dashed line at 1h/core is added for readability. The bottom panel shows the sum of the computing time used for each setup, both for the total program runtime (brown) and for the two main modules together (pink).}
    \label{fig:RTNLTE_timings}
\end{figure}
Figure~\ref{fig:RTNLTE_timings} displays the computational efficiency of the \textsc{RadTrans} module and the \textsc{NLTE\_solver} module within the \texttt{ExTraSS} framework, \txtred{for} setups with an increasing number of nodes -- with each node having $512\,$GB of RAM and 128 cores. The starting point for each run, which computed the 3D model from \vBPIII26, was created by setting the $\gamma$-ray energy deposition, running the first \textsc{NLTE\_solver} and making a checkpoint there. This checkpoint was loaded at the start of each run, which then performed one iteration of the \textsc{RayTraceGrid} and \textsc{NLTE\_solver} modules (see Figure~\ref{fig:TotalExTraSS_flowchart}).

The top panel in Figure~\ref{fig:RTNLTE_timings} shows the computational fraction of the total time spent by \texttt{ExTraSS} for the \textsc{RayTraceGrid} module (green circles), the \textsc{NLTE\_solver} module (orange stars), the translation steps between them\footnote{Relatively minor scripts which e.g. convert the number of photons absorbed by photoionization and photoexcitation into rates that are used inside the matrix solver during the calculation of the level populations.} (purple triangles) and the core-idling time during the run (fuchsia diamonds). Idling time occurs when the first cores complete a module and have to wait for the last cores to catch up. In the middle panel of Figure~\ref{fig:RTNLTE_timings}, the violin plots show the spread of runtimes for the two modules for all cores, with the \textsc{RayTraceGrid} time in blue and the \textsc{NLTE\_solver} time in orange. The black dot in this panel indicates the total \texttt{ExTraSS} wall-clock runtime. In the bottom panel, the total CPU h usage of each setup is shown, indicating the total \texttt{ExTraSS} CPU cost (in brown) and for the runtime of the \textsc{NLTE\_solver} and \textsc{RayTraceGrid} modules together, discounting the idling time (in pink).

Figure~\ref{fig:RTNLTE_timings} shows that, as long as good load-balancing can be maintained, an overall computational efficiency of $\sim0.8$ can be achieved. The model used has $N_\phi=30$, and the domain decomposition employed by \texttt{ExTraSS} creates domains equal $\phi$-size, or as close to that as possible. The setups with 9 and 12 nodes cannot achieve equally sized domains for each node, and thus suffer in computational efficiency and see an increase in idling time. As more nodes are introduced, the overhead cost in the translation scripts also increases slightly.

The setups with 6, 9 and 12 nodes also find very similar contributions from the \textsc{NLTE\_solver} module to the total runtime, indicating that optimized load balancing is much more important for the \textsc{RayTraceGrid} module. The wall-clock time still decreases by adding more nodes beyond an ideal $\phi$-decomposed setup (see the black dots in the middle panel of Figure~\ref{fig:RTNLTE_timings} -- from $\sim70\,$minutes to $\sim40\,$minutes, going from 6 to 12 nodes), but the total CPU h usage does increase past 6 nodes. The computing time spent inside the \textsc{NLTE\_solver} is practically identical across the setups, but with more nodes there is more idling time, which does cause the \textsc{NLTE\_solver} to have a stronger CPU h contribution for setups with more nodes. This effect is even stronger for the \textsc{RayTraceGrid} module, where issues with load balancing are more pronounced. This is because the larger domains both have to generate more rays (thus increasing their computing time), but also receive more incoming rays (as they have a larger fraction of the edge of the grid).

The setups with 1, 3 and 6 nodes display kind of a bimodal distribution in the \textsc{RayTraceGrid} computation times per core. For the 1 node setup, this can likely be attributed to the specific model used; the radiative transfer calculations are more heavy on the southern side of the model (due to the presence of several strong $^{56}$Ni plumes, see \citealt{stockinger2020three} and \vBPIII26); with $N_\theta\times N_\phi = 15\times30$ and 127 worker cores this causes roughly half of the workers to be biased for computations in the northern half, thus finishing faster. For the setups with 3 and 6 nodes, this likely plays a role as well (but to a lesser degree), but here there is a minor difference as the viewing angles are set up for $20\times20$ ($N_{\theta\text{, VA}} \times N_{\phi\text{, VA}}$) by default, and this means that the domains do not have an equal distribution in how many rays will (attempt to) escape to the viewers. Therefore, the domain that has the largest fraction of the `viewers' will receive more rays from its neighbours, thus making that domain somewhat slower.

The efficiency comparison here was only done for the first iteration of the \textsc{RayTraceGrid} and \textsc{NLTE\_solver} modules; for subsequent iterations generally the trend is that the \textsc{NLTE\_solver} computations become somewhat faster (as \texttt{ExTraSS} approaches global convergence, fewer deviations in the level populations occur) while for the \textsc{RayTraceGrid} module \emph{usually} the computational time increases somewhat as generally the number of lines capable of photoexcitation increases, and computations to photoionization remain equally costly at subsequent iterations. However, the exact evolution of the runtimes with iterations depend on the model and on the load balancing -- models with optimized domain decomposition should achieve an efficiency of around 0.8 or even better.

\subsection{Global convergence} \label{ssec:globconv}
Within the \textsc{NLTE\_solver}, the goal is to achieve \emph{local} convergence, that is the temperature and level populations are balanced such that the total emission matches the total $\gamma$-ray deposition in that cell. As Figures~\ref{fig:SFNLTE_stab} and \ref{fig:EXCION_stab} show, local convergence is broadly achieved (with most setups at >0.995). However, on a global scale this has to be coupled to the radiative transport and the impacts of photoionization and photoexcitation on the level populations and temperature. The global convergence is checked by comparing the energy in the emergent spectra, integrated over all rays (plus all emission outside the stored range) against the global $\gamma$-ray energy deposition. Global convergence is reached \txtred{when} the ratio of bolometric luminosity to $\gamma$-ray deposition ($L_\text{bol}$ / $L_\text{dep}$) \txtred{stabilizes at a value close to} unity\footnote{Radiative transfer introduces some adiabatic losses in a homologous flow, so the typical value is in the range of 0.9--1.}.

\begin{figure}
    \centering
    \includegraphics[width=\linewidth]{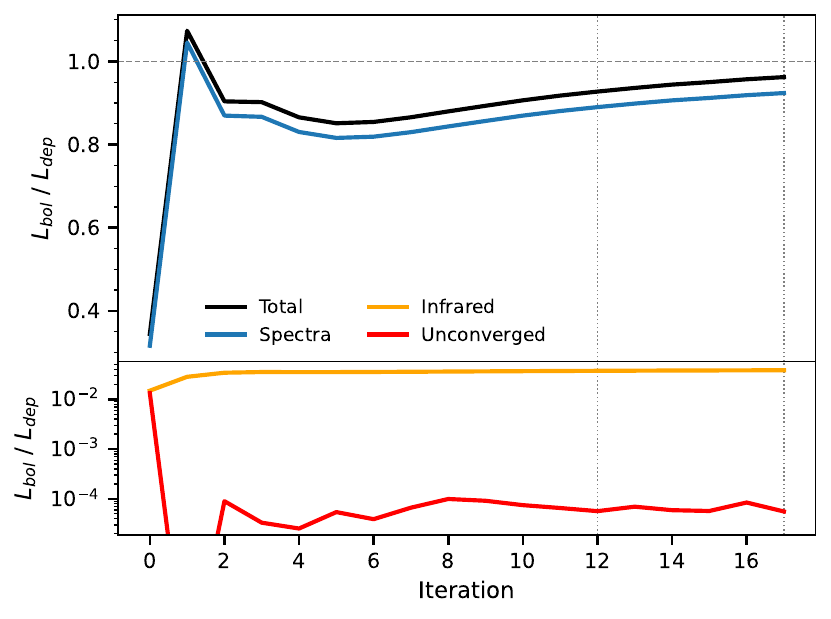}
    \caption{The global convergence per iteration for the model from \vBPIII26, separated by the total sum (black, top panel), emergent spectral contribution (blue, top panel), infrared contribution past the spectral range (orange, bottom panel) and the non-converged cells from \textsc{NLTE\_solver} (red, bottom panel). A gray dashed line is added at unity for visual aid, while the two dashed vertical lines indicate separate run iterations.}
    \label{fig:global_conv}
\end{figure}
An example of a typical global convergence curve is shown in Figure~\ref{fig:global_conv}, for the 3D model from \vBPIII26. The black curve shows the total convergence ratio, which is compromised of the three other components shown: the emergent spectra (in blue), the infrared contribution (emission beyond $25\,000\,\angstrom$ in this case; in orange) and the deposition in the non-converged cells (in red). The bottom panel is shown on a logarithmic scale to better display that past the first iteration, the contribution of non-converged cells is negligible, while the infrared contribution tops out at a few percent of the total. Generally, in the first few iterations the energy budget can fluctuate quite strongly (even overshooting in the case shown in Figure~\ref{fig:global_conv}) but then settles, and asymptotically rises towards unity. This specific run was done with four nodes on Dardel (i.e. 512 cores) in two separate batches, with the first one covering iterations 0 through 12, and the second one iterations 13 through 17. Loading the solution of iteration 12 as starting point for iteration 13 has no noticeable effect on the global convergence or the contribution of non-converged cells, which is an important check that such sequenced running (sometimes needed due to time limits for individual jobs) works as intended. After the 17th iteration, the total energy balance has reached 0.962, and with three consecutive iterations above 0.95, the model is considered converged.

\section{\txtred{Code Validation}} \label{sec:validation}
\begin{table*}
\centering
\begin{tabular}{l l l l l}
\hline
  & \multicolumn{2}{c}{\texttt{ExTraSS}} & \multicolumn{2}{c}{\texttt{SUMO}} \\  
\hline
\hline
Excitation  & Max 100 per ion & Calculated alongside ionization & Max 500 per ion & Calculated per ion \\
            & Collisional (de-)excitation & Spontaneous de-excitation & Collisional (de-)excitation & Spontaneous de-excitation \\
                & Photoexcitation & Non-thermal excitation & Photoexcitation & Non-thermal excitation \\
\hline
Ionization  & Photoionization for $n_i\leqslant25$ & Calculated alongside excitation & Photoionization for $n_i\leqslant50$ & Calculated per ion  \\
            & Non-thermal ionization &    & Non-thermal ionization & Collisional ionization  \\
            &     &    & Penning ionization & \\
\hline
Emission mechanisms & Lines &  Recombination continuum &  Lines &  Recombination continuum \\
            & H, He 2-photon &  & H, He 2-photon & Bremsstrahlung \\
\hline
Transfer effects & \multicolumn{2}{l}{Photoionization, calculation until $\lambda_\text{PI-edge}$} & \multicolumn{2}{l}{Photoionization, photons stop at $\lambda_\text{PI-edge}$}  \\
            & \multicolumn{2}{l}{Line-by-line photoexcitation, $\tau_\text{Sobolev}$}  & \multicolumn{2}{l}{Line-by-line photoexcitation, modified $\tau_\text{Sobolev}$}  \\
            & \multicolumn{2}{l}{}  &  Ly$_\alpha$, Ly$_\beta$-pumping &  e$^-$ scattering  \\
            &  &  &  Free-free absorption &  \\
\hline
\end{tabular}
\caption{\txtred{A comprehensive overview of the microphysics employed by \texttt{ExTraSS} and \texttt{SUMO} for the \J18 model. The excitation and ionization components are listed, as well as the different emission mechanisms and radiative transfer effects.}}
\label{tab:model_setups_micro}
\end{table*}
When validating our results from the radiative transport, we compare \texttt{ExTraSS} against 1D results from \texttt{SUMO} through several models. Firstly, the 1D results come from the model analysed by \citet[][\J18 hereafter]{jerkstrand2018emission}, and as such here we computed a 1D-turned-3D version thereof with \texttt{ExTraSS}. Additionally, a 3D explosion model of the same progenitor star is also available \citep{stockinger2020three} and this 3D explosion is investigated in more detail in \vBPIII26. Here we also consider the results from this full 3D model as extra basis to compare against and to validate the radiative transfer module. The comparison between the 1D and 1D-turned-3D model gives us the most direct handle to compare \txtred{physical conditions and spectral outputs,} photoionization and photoexcitation rates\txtred{,} and level population distributions between \texttt{ExTraSS} and \texttt{SUMO} to confirm that \texttt{ExTraSS} converges correctly. \txtred{In Table~\ref{tab:model_setups_micro}, we outline the microphysics and approximations used in \texttt{ExTraSS} and \texttt{SUMO} for the \J18 setup.}

\txtred{We also compare \texttt{ExTraSS} against the work by \citet{blondin2022standart}, in particular their `toy06' model, as that has previously been used to compare nebular phase codes.}

\subsection{\txtred{Temperature, ionization and spectra}} \label{ssec:codecompare}
\txtred{First}, we compare how \texttt{ExTraSS} spectral outputs compare against previous 1D work. For this we compare both against \J18 with our 1D-turned-3D setup (at 400 days), as well as the `toy06' model of \citet{blondin2022standart}, at 200 days. \txtred{For these comparisons, we look at both the physical conditions as well as the spectral outputs. A deeper investigation into the radiation field is given in Section~\ref{ssec:radtrans_out}.}

\subsubsection{\J18 in 3D}
The progenitor star used in the companion paper (\vBPIII26) was previously exploded and analysed in 1D with \texttt{SUMO} by \J18. The explosion energy, nuclear network and ejecta masses (slightly) vary between 1D and 3D, preventing a direct one-to-one comparison. By using the 1D-turned-3D model instead, we can make a more direct one-to-one comparison. To further minimize the differences between the setups, also the $\gamma$-ray deposition values from \J18 were used\footnote{The scattering code used for the $\gamma$-ray deposition in \texttt{ExTraSS} is technically more accurate, but for a more direct code comparison we chose to use the 1D data instead.}, and only the \textsc{NLTE\_solver} and \textsc{RayTraceGrid} computations from \texttt{ExTraSS} were employed. Although \J18 computed models at 200, 400 and 600 days, here we only compare at 400 days.

\begin{figure}
    \centering
    \includegraphics[width=\linewidth]{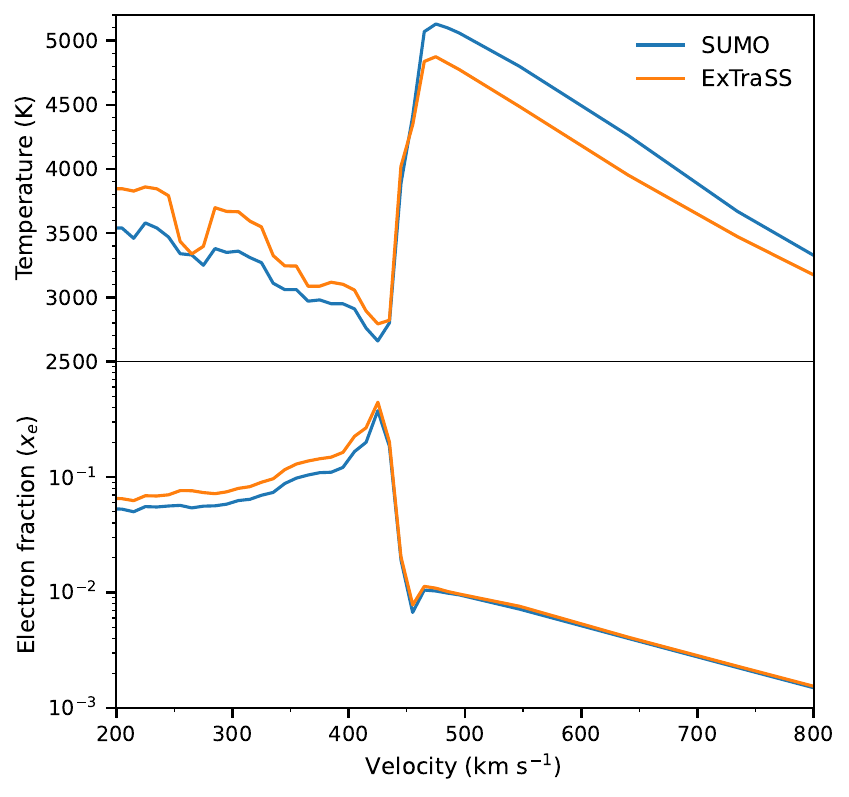}
    \caption{The temperature (top) and free electron fraction ($x_e$, bottom) at 400 days for the model, computed with \texttt{SUMO} (blue) and \texttt{ExTraSS} (orange).}
    \label{fig:J18-TXE_compare}
\end{figure}

\textbf{Physical conditions.}
In Figure~\ref{fig:J18-TXE_compare} we plot the temperature (top panel) and free electron fraction ($x_e$, bottom panel) for \texttt{SUMO} (in blue) and \texttt{ExTraSS} (in orange). The agreement is overall good. For the temperature, the innermost ejecta are somewhat hotter when computed with \texttt{ExTraSS} while the temperatures in the envelope are somewhat higher for \texttt{SUMO}. The differences in $x_e$ are \txtred{also small with \texttt{ExTraSS} giving slightly higher $x_e$ in the inner ejecta}, with both codes computing very similar \txtred{fractions in the envelope}.  

\textbf{Spectral output.}
\begin{figure*}
    \centering
    \includegraphics[width=.98\linewidth]{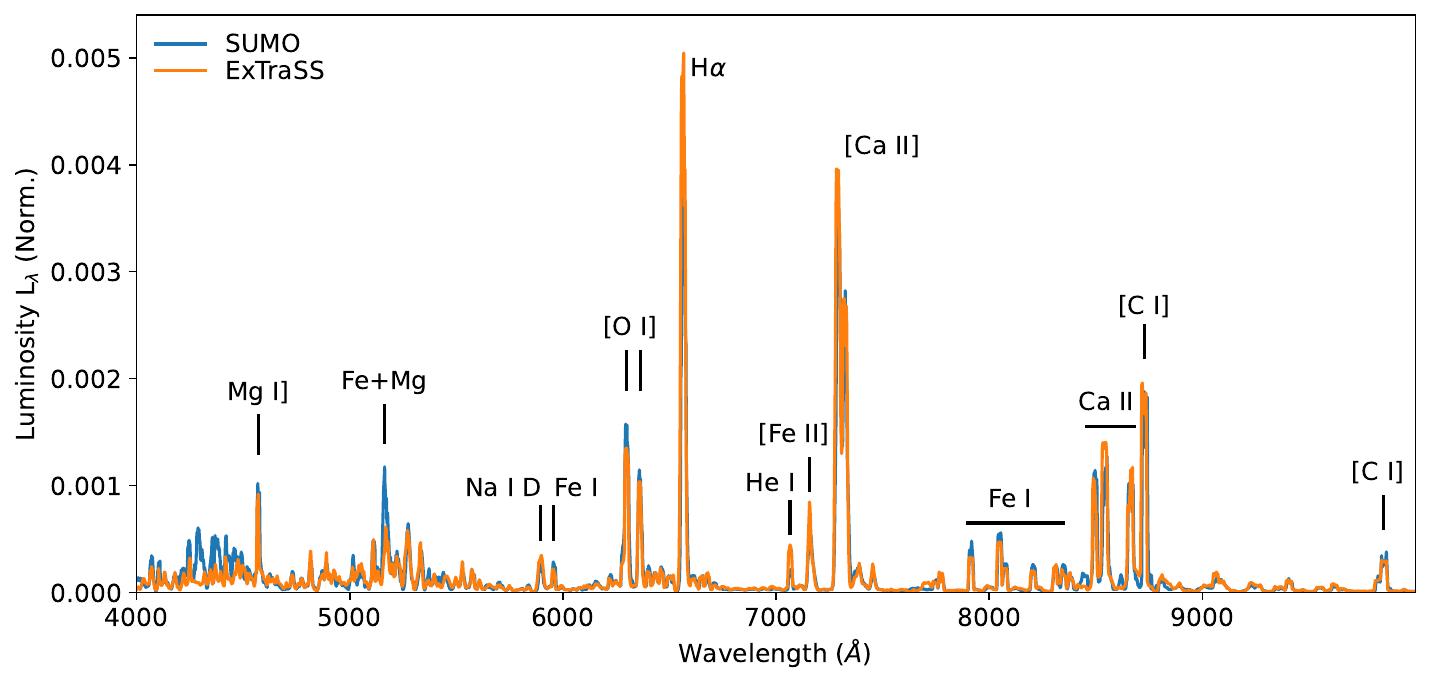}
    \caption{Comparison between outputs of \texttt{SUMO} (blue) and \texttt{ExTraSS} (orange) for the (1D) ejecta model of \J18, at 400 days. \texttt{ExTraSS} used the same $\gamma$-ray deposition values as the \texttt{SUMO} model to avoid differences arising from this. The main emission lines are marked.}
    \label{fig:J18_3Dcomp}
\end{figure*}
In Figure~\ref{fig:J18_3Dcomp}, the spectra between \texttt{SUMO} (1D, blue) and \texttt{ExTraSS} (3D, orange) are compared for the \J18 model at 400 days. The overall agreement between the two outputs is good, although various minor differences appear.

One of the more noticeable differences is that \texttt{SUMO} predicts slightly more emission in the $4000$--$4500\,\angstrom$ range, largely from \txtred{Fe~I and Fe~II}. \texttt{ExTraSS} instead has some additional H$\alpha$ emission. \texttt{SUMO} predicted a noticeable feature at $5180\,\angstrom$, which is a blend of Mg~I $\lambda\,5180$, Fe~I and Fe~II features, which is somewhat less strong in \texttt{ExTraSS}.

For practically all other clearly identifiable features -- see the markings in Figure~\ref{fig:J18_3Dcomp}, plus [O~I] $\lambda\,5577$ and O~I $\lambda\,7774$ -- the two codes find very similar line luminosities. \texttt{SUMO} has a specific treatment for Ly$\beta$ resonance pumping of the O~I $\lambda\,8446$ line, which is not included in \texttt{ExTraSS}, and this creates a noticeable but small difference.

Quite a lot of the spectrally noticeable differences between the two codes comes from deviations in the Fe emission -- \texttt{ExTraSS} gets stronger [Fe~II] $\lambda\,7155$, while in \texttt{SUMO} there is more Fe emission in the range of $4000$--$4500\,\angstrom$. In \texttt{SUMO}, most of the emergent spectrum in the $4000$--$4500\,\angstrom$ range originates in the H-zone (see Figure~9 in \J18), and with \texttt{ExTraSS} we find somewhat stronger H$\alpha$ emission, which might be appearing instead of the Fe emission. One difference which remains between the codes is the number of excited states included per ion, which is capped at 100 for \texttt{ExTraSS} while in \texttt{SUMO} Fe~I has 496 levels, and Fe~II has 533.

\subsubsection{Model toy06 from \citet{blondin2022standart}}
The code comparison by \citet{blondin2022standart} considered 10 codes, of which three were used to compute nebular phase spectra (\texttt{ARTIS-nebular} \citep{shingles2020monte}, \texttt{CMFGEN} \citep{hillier2012time}, and \texttt{SUMO} \citep{jerkstrand201144ti,jerkstrand2012progenitor}) and four models, of which one was used for the nebular phase analysis -- `toy06', at 200 days. This model has $M_{\mathrm{ej}} = 1.0\,M_\odot$ of which $M(^{56}\mathrm{Ni})=0.6\,M_\odot$, $E_\mathrm{kin}=10^{51}\,$erg and consists of 807 spherical shells with width $\Delta v=50\,\mathrm{km}\,\mathrm{s}^{-1}$ with an exponential density profile \citep[see also][for more details]{blondin2022standart}\footnote{Figures~\ref{fig:toy06_physical} and \ref{fig:toy06_spectra} were generated using the publicly available Github: \hyperlink{https://github.com/sn-rad-trans/data1}{https://github.com/sn-rad-trans/data1}}. For the computations with \texttt{ExTraSS}, the ejecta were resampled to \kms{\Delta v=500} and truncated at \kms{v=30\,000}, akin to how \texttt{SUMO} processed the model.

\begin{figure}
    \centering
    \includegraphics[width=\linewidth]{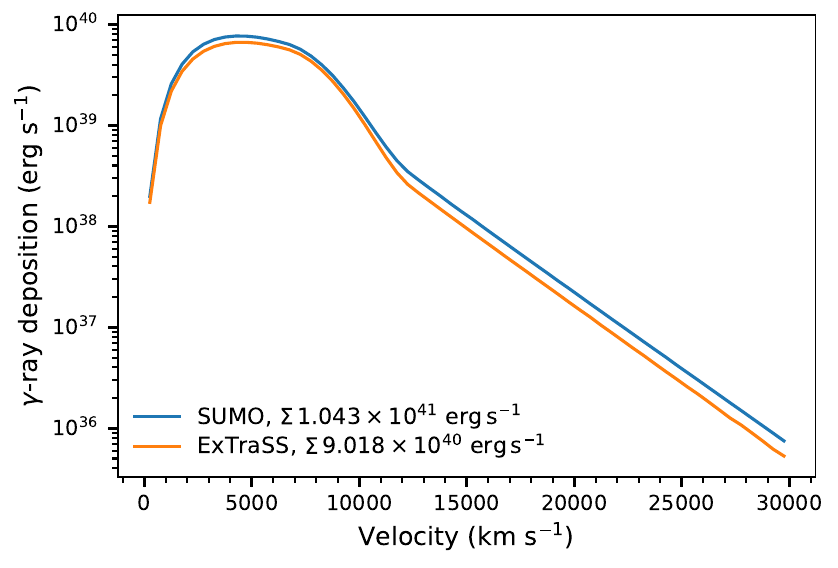}
    \caption{\txtred{The $\gamma$-ray deposition rates of \texttt{SUMO} (blue) and \texttt{ExTraSS} (orange) for the `toy06' model from \citet{blondin2022standart}, at 200 days.}}
    \label{fig:toy06_deposition}
\end{figure}

\begin{figure*}
    \centering
    \includegraphics[width=.495\linewidth]{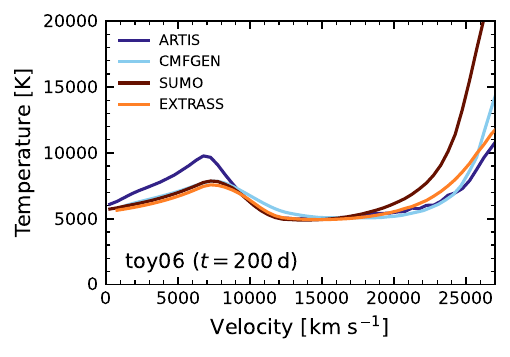}
    \includegraphics[width=.495\linewidth]{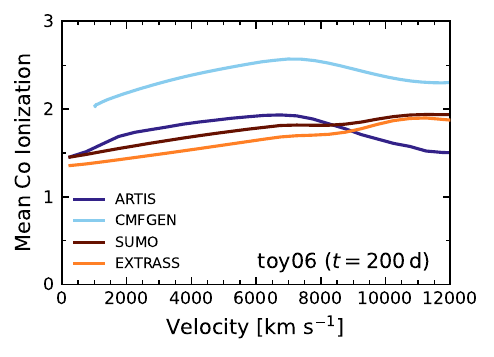}
    \caption{The temperatures (left) and mean Co ionization fractions (right) for the three nebular phase codes from \citet{blondin2022standart}, alongside \texttt{ExTraSS} (in orange). The Co panel zooms in on the ejecta up to \kms{12\,000}, as there is no Co at higher velocities. The `toy06' model, as well as the other codes, are all described in detail in \citet{blondin2022standart}.}
    \label{fig:toy06_physical}
\end{figure*}

\begin{figure*}
    \centering
    \includegraphics[width=.98\linewidth]{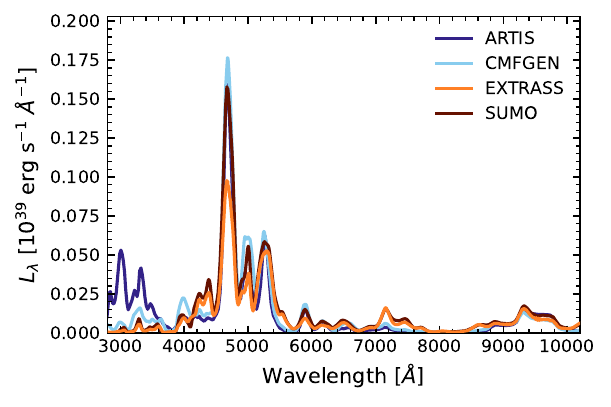}
    \caption{Spectral comparison at 200 days for the toy06 model from \citet{blondin2022standart}, for the three nebular phase codes compared there, and \texttt{ExTraSS} (in orange) which is presented in this work. The `toy06' model, as well as the other codes, are all described in detail in \citet{blondin2022standart}.}
    \label{fig:toy06_spectra}
\end{figure*}

\txtred{\textbf{Energy deposition.} Since the comparison with \J18 was done with equal $\gamma$-ray deposition already, for the comparison with `toy06', we also used the $\gamma$-ray deposition from \texttt{ExTraSS} itself rather than using \texttt{SUMO} data again. In Figure~\ref{fig:toy06_deposition} the comparison between the energy deposition against velocity is shown, for \texttt{SUMO} (blue) and \texttt{ExTraSS} (orange). The other nebular phase codes are extremely similar to \texttt{SUMO} and thus not shown. Overall, the energy deposition with \texttt{ExTraSS} is $\sim10\,\%$ lower, which is consistent throughout the ejecta. \texttt{SUMO} uses a gray opacity for the $\gamma$-rays, while with \texttt{ExTraSS} the Compton scattering formalism is used (see \Jerk20 for details).}

\textbf{Physical conditions.} In \citet{blondin2022standart}, profiles for temperature and mean Co ionization fraction were compared. In Figure~\ref{fig:toy06_physical}, we display these from the three codes used in the comparison work, alongside the ones obtained by \texttt{ExTraSS} (in orange).

Compared to the other codes, for the innermost ejecta \kms{v<9000}, \texttt{ExTraSS} computes the lowest temperatures, but still relatively close to in particularly the \texttt{CMFGEN} and \texttt{SUMO} profiles. At velocities up to \kms{\sim20\,000}, the temperature in \texttt{ExTraSS} is very similar to all other codes, and at the highest velocities it falls in between the others, with an increase which begins early like for \texttt{SUMO}, but which proceeds more slowly, more akin to \texttt{ARTIS-nebular}.

For the mean Co ionization, \texttt{ExTraSS} computes somewhat lower ionization fractions up to \kms{8000}, after which it becomes very similar to \texttt{SUMO}, and the \txtred{values} are higher than for \texttt{ARTIS-nebular}, although they are lower than \texttt{CMFGEN} at all velocities. Generally, the somewhat lower Co ionization might be linked to the somewhat lower temperatures for the inner side of the ejecta\txtred{, and the somewhat lower energy deposition (see Figure~\ref{fig:toy06_deposition})}.

\textbf{Spectral output.} In Figure~\ref{fig:toy06_spectra}, we show the comparison for the spectral output. In general, \texttt{ExTraSS} makes similar predictions as the other codes, in particular compared to \texttt{SUMO}. The two main areas of discrepancy are at $\sim4600$--$5000\,\angstrom$, where the other codes find relatively similar peak luminosities for the Fe emission, but which is noticeably lower in \texttt{ExTraSS}. For the peak at $\sim4600\,\angstrom$, the peak luminosity is about $40\,\%$ lower, while for the line at $\sim5000\,\angstrom$ \texttt{ExTraSS} predicts about $30\,\%$ less luminosity. Elsewhere, the spectral prediction agrees well at least with \texttt{SUMO} (e.g. for the Fe feature around $\sim7200\,\angstrom$) or even all three codes ($5100-6400\,\angstrom$, and past $\sim8800\,\angstrom$). There is \txtred{some marginal} Co emission around $\sim8000\,\angstrom$ which comes from Co~I, which could perhaps be explained by the somewhat lower Co ionization fraction in \texttt{ExTraSS}.

\texttt{ExTraSS} uses the same atomic data as \texttt{SUMO} with the exception of the number of levels per ion, which is capped at 100 for \texttt{ExTraSS}. In particular for Fe and Co, this could be a reason why some (small) differences appear, although a difference such as for the Fe feature at $\sim4600\,\angstrom$ might be too great to be caused solely by that.

\subsection{\txtred{Radiation field rates}}  \label{ssec:radtrans_out}
\txtred{Here, we will take a more detailed look at the rates computed for photoionization and photoexcitation by \texttt{ExTraSS} and \texttt{SUMO}, for both the models considered above.}

\subsubsection{Photoionization}  \label{ssec:PI_valid}
As described in Section~\ref{sssec:PI_RT}, the radiative transfer module computes photoionization and photoexcitation rates. \txtred{First}, we compare the photoionization rates for the 1D model from \J18, \txtred{the} 3D version thereof (see also Section~\ref{ssec:codecompare}), and the 3D s9.0 model from \vBPIII26. The 1D rates are computed with \texttt{SUMO}, while the 3D setups are computed with \texttt{ExTraSS}.

\begin{figure}
    \centering
    \includegraphics[width=\linewidth]{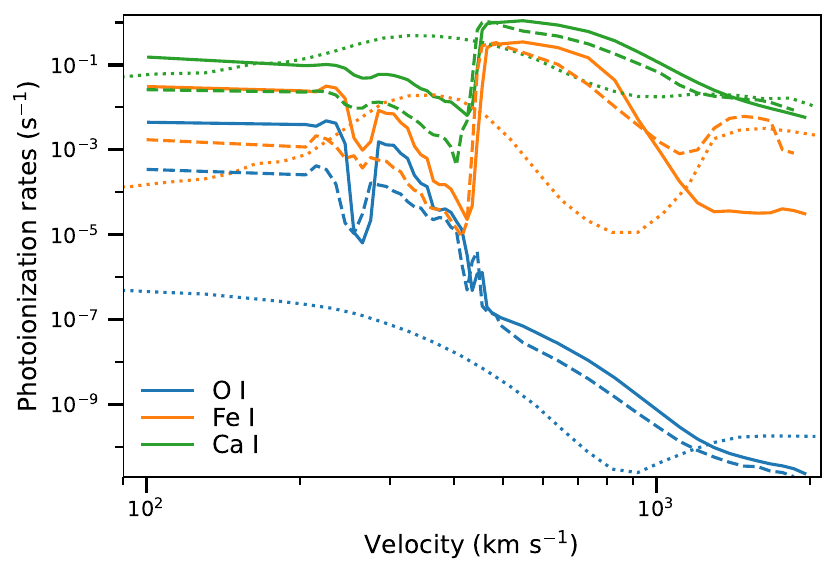}
    \caption{A comparison between the photoionization rates of \texttt{ExTraSS}, for the s9.0 model from \vBPIII26 (dotted lines) and the 3D version (solid lines) of \J18, with the 1D rates from \texttt{SUMO} (dashed), all at 400 days. Photoionization rates for O~I to O~II are shown in blue, for Fe~I to Fe~II in orange, and for Ca~I to Ca~II in green.}
    \label{fig:PIrates_1D3D}
\end{figure}

In Figure~\ref{fig:PIrates_1D3D} the (angle-averaged, for the 3D models) photoionization rates for these three models are shown, with \texttt{ExTraSS} as solid lines (for the 3D version of \J18) and dotted lines (the model from \vBPIII26, a 3D explosion by \citealt{stockinger2020three} of the same progenitor star). The 1D rates of \texttt{SUMO} are shown as dashed lines to compare. For each model, we display the photoionization rates of neutral oxygen (blue), iron (orange) and calcium (green).

For the innermost ejecta (\kms{v<450}) in the \J18 model, the rates in \texttt{ExTraSS} are \txtred{at least} an order of magnitude larger. Towards the `bump' at \kms{450} (present due to the stratification of the ejecta), the rates for O and Fe become more similar. In the \J18 model, most mass resides at \kms{v>500}, where the rates for O and Ca are reasonably similar between the two codes, although \texttt{ExTraSS} has higher rates until deep into the envelope. For Fe, the rates in \texttt{ExTraSS} keep decreasing while \texttt{SUMO} sees an increase at \kms{v\gtrsim1000}.

Compared to the 3D explosion model from \vBPIII26, which has lower expansion velocities and thus higher densities, very different patterns appear. For O, the rates never exceed $10^{-6}\,$s$^{-1}$, which is effectively also the highest the rate gets in the envelope in the 1D setup. The lowest rates occur not at the highest velocities but slightly before, at \kms{v\approx900}, before increasing towards the outermost zones. Generally, the range of values computed for the photoionization rates in the full 3D setup are lower due to the lower densities and more mixed ejecta, but within the same range as the \J18 model.

The photoionization rate for Fe is also lower in the \vBPIII26 model, peaking at \kms{v\approx300} and towards the outer envelope, similar to oxygen. Due to the lack of stratification, there is no strong peak in the Fe photoionization rates, and instead it ranges from $10^{-2}$ to $10^{-5}\,$s$^{-1}$. The photoionization rates for Ca generally follow the same trend as Fe in \vBPIII26, but two to three orders of magnitude larger.

\begin{figure}
    \centering
    \includegraphics[width=\linewidth]{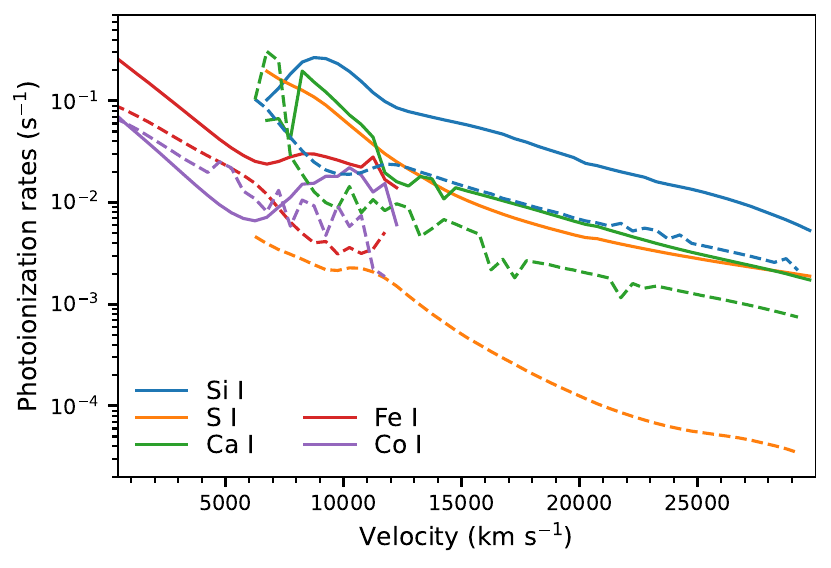}
    \caption{\txtred{The photoionization rates of \texttt{ExTraSS} (solid lines) and \texttt{SUMO} (dashed), for the `toy06' model from \citet{blondin2022standart}, at 200 days. Photoionization rates for Si~I to Si~II are shown in blue, for S~I to S~II in orange, for Ca~I to Ca~II in green, for Fe~I to Fe~II in red, and Co~I to Co~II in purple. Fe and Co are only present up to \kms{12\,000}, while Si, S and Ca are only present from \kms{6000} onwards (see also \citealt{blondin2022standart}).}}
    \label{fig:PIrates_toy06}
\end{figure}

\txtred{In Figure~\ref{fig:PIrates_toy06} the photoionization rates are compared between \texttt{ExTraSS} and \texttt{SUMO} for the `toy06' model from \citet{blondin2022standart}, at 200 days. For Fe and Co, \texttt{SUMO} computes very similar rates, while \texttt{ExTraSS} has somewhat higher Fe photoionization rates and lower rates for Co, although the differences between the two codes are only a factor of $\sim3$ at most. For Ca, the photoionization rate in \texttt{ExTraSS} is roughly a factor 2 higher across all velocities, while for Si it is a factor $\sim3$ again. The gap in the rates for S~I are much bigger, with \texttt{ExTraSS} computing a rate at least one order of magnitude larger.}

\txtred{The differences in the photoionization rates between the two codes are larger for this model than in the comparison for the s9.0 model from \J18, although they only become particularly big for S. As the energy deposition in these models is different (while in the \J18 comparison the $\gamma$-ray deposition was equal) larger differences are expected to arise, as the available energy budget will also lead to differences in non-thermal ionization rates. As the right panel of Figure~\ref{fig:toy06_physical} shows, the mean Co ionization fraction in \texttt{ExTraSS} is always slightly lower, even in the regions where the photoionization rate is higher, which originates with different non-thermal ionization rates which the photoionization rates is partially compensating for.}

\subsubsection{Photoexcitation}  \label{ssec:PE_valid}
\txtred{The photoexcitation rates will first be compared for the \J18 model, the 3D version thereof, and s9.0 from \vBPIII26 (as in Section~\ref{ssec:PI_valid})}. The 1D rates are computed with \texttt{SUMO}, while the 3D setups are computed with \texttt{ExTraSS}. \txtred{Secondly, we also show the photoexcitation rates for the `toy06' model for \texttt{ExTraSS} and \texttt{SUMO}.}

\begin{figure}
    \centering
    \includegraphics[width=\linewidth]{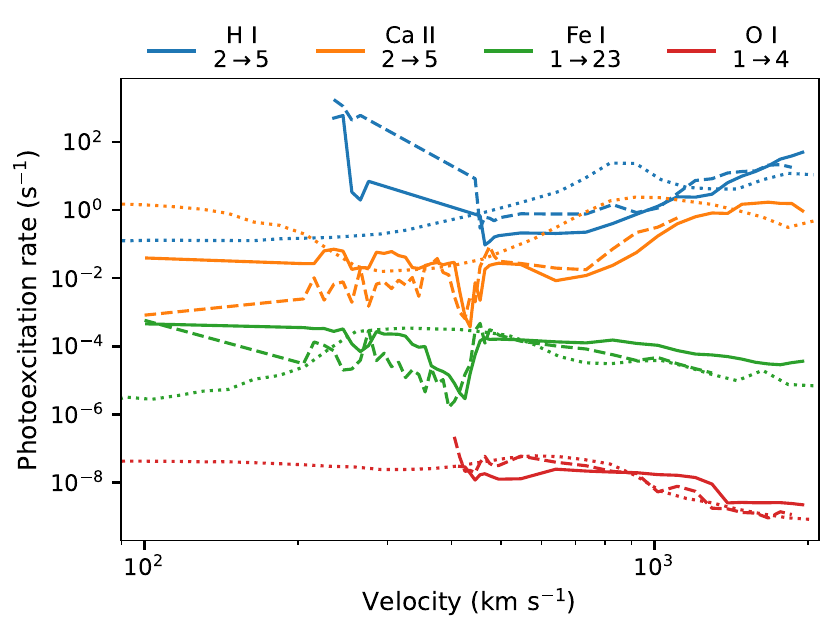}
    \caption{A comparison between the photoexcitation rates of \texttt{ExTraSS}, for the model from \vBPIII26 (dotted lines) and the 3D version (angle-averaged, solid lines) of \J18, with the 1D rates from \texttt{SUMO} (dashed), all at 400 days. H$\alpha$ is shown in blue, Ca~II $\lambda\,8498$ in orange, Fe~I $\lambda\,5060$ in green and [O~I] $\lambda\,6300$ in red.}
    \label{fig:PErates_1D3D}
\end{figure}

In Figure~\ref{fig:PErates_1D3D}, the three setups \txtred{of the \J18 model} are compared for four different elements: H$\alpha$ (transition $2\rightarrow5$, which is the bluest component in H$\alpha$), Ca~II $\lambda\,8498$ (the blue line in the NIR triplet), Fe~I $\lambda\,5060$ (the first allowed transition to the ground state in neutral Fe) and [O~I] $\lambda\,6300$ (the blue component of the [O~I] $\lambda\lambda\,6300,\,6364$ doublet). The solid and dashed lines have the same model basis (\J18), with the dashed computed with \texttt{SUMO} in 1D, and the solid with \texttt{ExTraSS} in 3D. The dotted lines are the rates from \vBPIII26 (a 3D model). The \texttt{ExTraSS} results show the angle-averaged rates.

For H$\alpha$, noticeable differences exist at \kms{200-400} in the \J18 model (\texttt{SUMO} as dashed lines, \texttt{ExTraSS} as solid lines), although H has a very low mass fraction at these velocities. The rates are more similar at higher velocities, in particular past \kms{v>800}. The denser model of \vBPIII26 finds rates of comparable magnitudes but with a peak at \kms{800}, which occurs in the part of the grid which has very low energy deposition but is still relatively dense.

The rates for Ca~II $\lambda\,8498$ differ significantly between the three setups at the lowest velocities \kms{<250}, but this region contains very little mass (less than $0.05\,M_\odot$ even in the denser \vBPIII26 model). At all other velocities, the rates between \texttt{ExTraSS} (solid) and \texttt{SUMO} (dashed) are in good agreement, although in the outer envelope (past \kms{1000}) for \texttt{SUMO} this line is too optically thin to compute photoexcitation rates, while for \texttt{ExTraSS} this occurs slightly further out. As with H$\alpha$, the model of \vBPIII26 finds a bump in the rates at lower velocities, but this peak rate is similar to the other setups.

For Fe~I $\lambda\,5060$, \texttt{ExTraSS} (solid) and \texttt{SUMO} (dashed) find very similar rates past \kms{400}, with \texttt{ExTraSS} giving somewhat higher rates at lower velocities. The sharp jump at \kms{\sim450} corresponds to kind of a `wall' between the Ni-rich core material and the He-envelope, where a significant fraction of the total energy deposition resides in the \J18 model. At higher velocities, the two codes are in excellent agreement for this specific rate. The \vBPIII26 model (dotted line) finds very low rates at the lowest velocities, and does not have a strict `wall' like the 1D setup, but transitions into these relatively higher rates at around \kms{250}. As with the other lines, the range of photoexcitation rates in the \vBPIII26 model are compatible with \J18.

For [O~I] $\lambda\,6300$, the line is not optically thick until \kms{400} in \J18, and generally \texttt{ExTraSS} (solid) and \texttt{SUMO} find similar rates, although \texttt{SUMO} computes somewhat higher rates at intermediate velocities and \texttt{ExTraSS} at the higher velocities. The model of \vBPIII26 has very similar rates as the other setup for the [O~I] line here, even at the lowest velocities, due to the mixing of the envelope material that contains oxygen.

The differences for the photoexcitation rates between \texttt{SUMO} and \texttt{ExTraSS} in the \J18 model are generally smaller than they were for the photoionization rates. This might be because in \texttt{ExTraSS}, the rays do not stop at the exact $\lambda_\text{PI}$, \txtred{although only the travel distance where the ray could photoionize is computed}. In \texttt{SUMO}, photons do stop at this `edge' of $\lambda_\text{PI}$\txtred{, so there is some subtle difference between the two codes}. For photoexcitation, both codes have the ray/photon stop at the $\lambda_\text{line}$ intersection\txtred{, which might explain why the differences are smaller for photoexcitation than photoionization}.

\begin{figure}
    \centering
    \includegraphics[width=\linewidth]{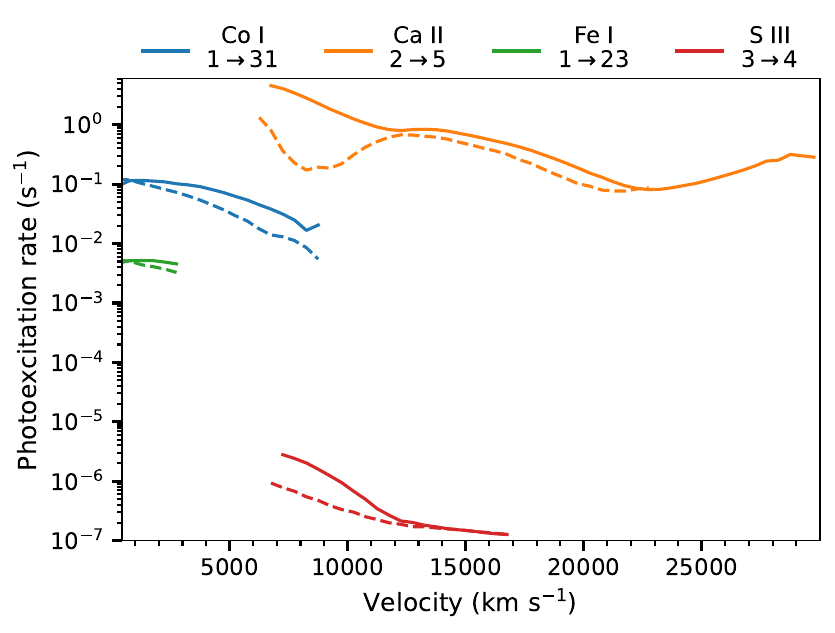}
    \caption{\txtred{The photoexcitation rates of \texttt{ExTraSS} (solid lines) and \texttt{SUMO} (dashed), for the `toy06' model from \citet{blondin2022standart}, at 200 days. Co~I $\lambda\,4234$ is shown in blue, Ca~II $\lambda\,8498$ in orange, Fe~I $\lambda\,5060$ in green and S~III $\lambda\,9531$ in red. Fe and Co are only present up to \kms{12\,000}, while S and Ca are only present from \kms{6000} onwards (see also \citealt{blondin2022standart}).}}
    \label{fig:PErates_toy06}
\end{figure}

\txtred{In Figure~\ref{fig:PErates_toy06}, the photoexcitation rates for \texttt{ExTraSS} (solid lines) and \texttt{SUMO} (dashed lines) are shown, for the `toy06' model of \citet{blondin2022standart} at 200 days. As this model contains fewer atomic species, we show Co~I $\lambda\,4234$ (instead of H$\alpha$, in blue) and S~III $\lambda\,9531$ (instead of [O~I] $\lambda\,6300$, in red). Generally, the codes find similar rates although various differences appear. For Co~I $\lambda\,4234$, the rates are initially very similar but \texttt{ExTraSS} computes higher rates towards higher velocities, and past \kms{9000} the line becomes optically thin in for codes. For Fe~I $\lambda\,5060$, the transition to optically thin occurs even earlier, at \kms{2500}, however for the small part where the codes compute photoexcitation they find good agreement. Ca~II $\lambda\,8498$ is only present from \kms{6000}, and initially \texttt{ExTraSS} computes rates approximately an order of magnitude larger. Further out there is good agreement yet it should be noted that this line becomes optically thin in \texttt{SUMO} at \kms{23\,000}, but remains above the $\tau_\text{line}>10^{-4}$ criterion in \texttt{ExTraSS}. Lastly, S~III $\lambda\,9531$ also appears at \kms{6000} and is present until \kms{17\,000} for both, with \texttt{ExTraSS} initially at rates a factor 3 higher, but equal rates from \kms{12\,000} onwards.}

\txtred{We don't show any rates of Si, as there are no transitions in the optical range that have an optical depth above $10^{-4}$, and thus photoexcitation does not take place for these transitions. For S, we chose S~III $\lambda\,9531$ as this generates some noticeable emission (it is part of the bump seen in Figure~\ref{fig:toy06_spectra}) and also for S there are no transitions in the optical that consistently have an optical depth above $10^{-4}$. For Ca and Fe we chose to show the same lines as when comparing the \J18 model, even though for in particular for Fe there are other lines we could have shown. Several of these lines have photoexcitation rates where \texttt{ExTraSS} is a factor $2-3$ higher, but the general `flow' of the rates is the same in both codes.}

\txtred{As with the \J18 model, generally the photoexcitation rates in the `toy06' model agree better than the photoionization rates, with most lines computing rates within a factor of 3, and even smaller differences for large parts. Areas where more substantial differences appear tend to be regions where the mass fraction of those elements becomes small, which can cause relatively small changes in level population have much stronger effects on the optical depth of lines and thereby the impact on the photoexcitation.}

\subsection{Level populations}  \label{ssec:levelpops}
\txtred{The level population solvers combine the photoionization and photoexcitation rates with the various other excitation and ionization processes that \texttt{ExTraSS} and \texttt{SUMO} account for (see also Table~\ref{tab:model_setups_micro}). The outcome of this are the level populations for all elements across all cells/zones in the codes. In Appendix~\ref{app:levelpop_compare} we take a detailed look at how the level populations for the \J18 model compare, and see how the somewhat different photoionization and photoexcitation rates result in different populations, which still generate a largely similar final spectrum.}


\section{Discussion} \label{sec:discussion}
In a 3D NLTE radiative transfer code, there are many bits and pieces which can be designed in various ways. Section~\ref{sec:performance} displayed a variety of the options we considered for \texttt{ExTraSS} while also outlining how the code scales to larger setups, how the internal computations work and \txtred{Section~\ref{sec:validation} showed how} these compare to outputs of other similar codes. Here, we will discuss the radiative transfer implementation and some alternative options, and also outline some future targets for code upgrades.

\subsection{Radiative transfer implementation}
Several alternative options exist when choosing how to perform the radiative transfer scheme as implemented in \texttt{ExTraSS}. Our chosen scheme, as outlined above, has one core per node (the manager) dedicated to manage all the communication for the photon packets leaving the domain and for the packets entering, and distributing these incoming packets to the other cores (the workers). There is some minor risk that the workers can outpace their manager, which would lead to significant `buffering' of the communication chain \citep[][]{brunner2009efficient}, but even on Dardel where there are 128 cores per node, this does not occur, and on systems with fewer cores per node this should thus also not be a concern.

The worker-manager setup employed by \texttt{ExTraSS} does have the advantage that offloading incoming rays to other workers can be achieved easily. In theory, when a worker updates the main core on its progress, it could also inform the manager on its own core, with that manager prioritizing incoming data bundles to give to cores that are closest to finishing. However, it does bring some minor communication cost, which as mentioned above could potentially be a risk towards the communication stability. Setups which cut out the manager, and straight up `link' cores across nodes that have the same internal rank (e.g. core 1 on node A to core 129 on node B) have a communicational advantage, but the drawback that offloading becomes very complex (and then requires additional MPI communication). Alternatively, through clever use of memory sharing -- effectively sharing the rays themselves -- it would be possible to minimize the MPI communication setup within each node, but it would (likely) require so much of the memory to be allocated as shared memory that it significantly slows down the computations. It also has issues with load balancing, since it makes offloading incoming rays much more complicated without specific internal MPI communication.

As such, the scheme we chose for \texttt{ExTraSS} might be the more complicated one in terms of MPI communication, since it requires the chain of `worker-to-manager-to-manager-to-worker' to transfer a photon packet from one domain to another. However, by making bundles of rays which are transferred together rather than doing the transfer for individual rays, and through the use of \emph{non-blocking} communication, \texttt{ExTraSS} can achieve good computational efficiency in the radiative transfer (see Figure~\ref{fig:RTNLTE_timings}) which is also stable with respect to global convergence (see Figure~\ref{fig:global_conv}).

\subsection{Future code updates and applications}  
In its current form, \texttt{ExTraSS} is set up as a domain decomposed 3D NLTE radiative transfer code, which accounts for both photoionization and photoexcitation using the Sobolev formalism. It currently does not treat either electron scattering, dust or molecules; each could conceivably have impacts at various epochs in the nebular phase. Electron scattering would be important for earlier times, when densities are somewhat higher \citep[see also][]{jerkstrand2017spectra}. Dust tends to be more important at later times, as the ejecta cool and dust as well as molecules can begin to form. 

Molecules are a complex topic within nebular phase SNe. They can form quite early (e.g. carbon monoxide was detected 112 days after explosion in SN~1987A, \citealt{oliva1987micron}, \citealt{spyromilio1988carbon}) but can can also become depleted onto dust over time \citep{wooden1993airborne}. Molecules have been detected in old remnants \citep[e.g. Cas A,][]{rho2024shockingly}, as well as at early times, such as recently in the nearby hydrogen-rich SN~2023ixf \citep{park2025near} as well as a few months after explosion for the Type Ic SN~2020oi \citep{rho2021near}. Efforts to include molecules for spectral modelling are so far limited to 1-zone or 1D modelling \citep{liu1995oxygen,liljegren2020carbon,liljegren2023molecular,mcleod2024carbon}. 

Other improvements aside from the ones mentioned above, which can be made for \texttt{ExTraSS} in the future, are to account for time dependence, to add additional processes to the rate equations (e.g. charge transfer and collisional ionization, which are not yet treated), or to include free-free absorption in the radiative transfer. Models which have been used so far (\vBPI23, \vBPII24, \vBPIII26), have not included odd-Z elements aside from the Co created by the radioactive decay of $^{56}$Ni. However, several of these elements (e.g. N, Na) are known to have non-negligible emission in the optical range, and these could be included in the future by correcting for the solar abundance ratio. In the 1D-turned-3D model of \J18 which we compared against here, these odd-Z elements were included, and we show that \texttt{ExTraSS} is capable of handling these extra elements. As \texttt{ExTraSS} assumes homologous expansion, the $^{56}$Ni-bubble effect \citep{gabler2021infancy} can be accounted for in the hydrodynamical modelling of the input models, but not in the homologous `fast-forwarding' afterwards.

\section{Conclusions} \label{sec:conclusion}
In this work, we present \texttt{ExTraSS} (EXplosive TRAnsient Spectral Synthesis) -- a 3D NLTE radiative transfer code for supernovae. The code has been upgraded to include photoionization and photoexcitation rates computed from the global radiation field, using a domain decomposition approach which enables the detailed microphysics needed for NLTE modelling also in 3D. 

As a result of the addition of radiative transfer, \texttt{ExTraSS} now iterates between solving for levels populations, temperature, and radiation field. Here we describe how we set up and optimized both the level population solver and the radiative transfer module. The level population solver now also includes non-thermal excitations (in addition to ionizations) from the Spencer-Fano calculation \citep{spencer1954energy,kozma1992gamma}, which mandated an adjustment to the solver to optimize the run time. The radiative transfer module required the introduction of domain decomposition due to memory limitations, and our choices surrounding the design of this are discussed.

The results of the new radiative transfer module are compared to the 1D s9.0 model analysed with \texttt{SUMO} by \J18, alongside a 1D-turned-3D version thereof computed with \texttt{ExTraSS}. Additionally, some comparisons are made against the full 3D model of \vBPIII26, who use a 3D explosion model of the same progenitor star. The rates for photoionization across several species is shown to be compatible between \texttt{ExTraSS} and \texttt{SUMO}, while photoexcitation rates between the two models are in excellent agreement. Some variation exists in the exact level populations between the two codes, yet they do not lead to particularly large differences in the spectral output or predicted temperatures.

Additionally, we compare \texttt{ExTraSS} with the `toy06' model from \citet{blondin2022standart} at 200 days (their `nebular phase' epoch) against \texttt{ARTIS}, \texttt{CMFGEN} and \texttt{SUMO}, and find that the physical outputs and spectra are in good agreement with the other codes for this model.

With the upgrade presented in this work, \texttt{ExTraSS} has been established as a domain decomposed, 3D NLTE radiative transfer spectral synthesis code aimed at the nebular phase. It currently has only been applied to core collapse supernova models at late times, but in the future the code can be used for a wide variety of input models and different explosive transient mechanisms, if certain additional physical components are included. Such expansions will be outlined in future works, but in its current form, \texttt{ExTraSS} can be applied to a large set of SN models, to study their spectral signatures, and better constrain explosion properties, ejecta masses, and composition.


\section*{Acknowledgements}

The authors would like to thank the referee for their comments and suggestions which helped improve the clarity significantly. The authors acknowledge support from the European Research Council (ERC) under the European Union’s Horizon 2020 Research and Innovation Programme (ERC Starting Grant No. [803189], PI: Jerkstrand) and the Knut and Alice Wallenberg Foundation (Project ``Gravity Meets Light'', PI: Rosswog and Jerkstrand). The computations were enabled by resources provided by the National Academic Infrastructure for Supercomputing in Sweden (NAISS) at the PDC Center for High Performance Computing, KTH Royal Institute of Technology, partially funded by the Swedish Research Council through grant agreement no. 2018-05973. The computations in this work were performed on the CPU partition of the Dardel HPE Cray EX system, which is predominantly funded by the National Academic Infrastructure for Supercomputing in Sweden (NAISS).

\section*{Data Availability}

The Data underlying this article will be shared on reasonable request to the corresponding author. The \texttt{ExTraSS} data will be added to the publicly available Github (\hyperlink{https://github.com/sn-rad-trans/data1}{https://github.com/sn-rad-trans/data1}), following the format specified in \citet{blondin2022standart}.




\bibliographystyle{mnras}
\bibliography{bibliography}




\appendix

\section{Legacy code components}
In this Appendix, we describe some older code routines that have become deprecated since but were used in older works or which were upgraded into more detailed treatments.

\subsection{Optically thin spectra}   \label{ssec:thin_spectra}
From the converged cells, optically thin emission packets are generated which are subsequently summed together for each viewing angle, accounting for the Doppler-shifted wavelengths and fluxes. In the optically thin limit, the only radiative transfer effect to consider is the local Sobolev escape probability $\beta_S$ which is set by the optical depth $\tau_S$ within this approximation:
\begin{equation}
    \beta_S = \dfrac{1-\rm{e}^{-\tau_S}}{\tau_S},
    \label{eq:beta_Sobolev}
\end{equation}
\begin{equation}
    \tau_S = \dfrac{A_{ul}}{8\pi}\dfrac{g_u}{g_l}\lambda^3_{ul}n_l\bigg(1-\dfrac{g_l n_u}{g_u n_l} \bigg)t.
    \label{eq:tau_Sobolev}
\end{equation}
Here, the subscripts $u$ and $l$ refer to the upper and lower state, respectively, $A_{ul}$ is the transition strength from $u$ to $l$ with the corresponding wavelength $\lambda_{ul}$ for this transition. $g$ are the statistical weights, $n$ the number densities, and $t$ is the time.

The optically thin emission packets are generated by combining the Sobolev escape probability (Eq~\ref{eq:beta_Sobolev}) with the transition probability, the energy difference between $u$ and $l$, and the total amount of particles in the upper state, $N_u$, in that cell:
\begin{equation}
    \rm{Emission}_{ul}(\lambda) = A_{ul}\times\Delta E_{ul}\times\beta_{S,ul} \times N_u.
    \label{eq:thin_emissivity}
\end{equation}
By combining the emissivity from every transition from every element in every cell together, the full optically thin spectrum can be constructed, if the Doppler-shift effects on the observer's frame are accounted for to adjust the wavelength to the correct bin for the observer: $\lambda_{\rm{emit},ul}$ to $\lambda_{\rm{obs},ul}$. A similar correction has to be made for correcting this emissivity$_{ul}$ into the observer's frame. This completely optically thin emission is what was used in \vBPI23, and predominantly in \vBPII24, except for recombination emission with $\lambda>5000\,\angstrom$.

\subsection{On-the-spot photoionization}  \label{ssec:OTSPI}
For recombination with $\lambda\leqslant5000\,\angstrom$, \vBPII24 instead applied an approximate `on-the-spot' treatment for photoionization, as this effect mostly comes into play for blue wavelengths. The cut-off at $5000\,\angstrom$ was chosen as redder photons no longer had enough energy to do any photoionization, as only the ground state multiplet plus the next eight excited states were considered\footnote{Photoionization is much less effective at redder wavelengths, and as this approximation was only implemented in the emitting cell this limit had no significant impact on the total emission.}. 

In the work presented here, we have improved slightly on the recombination emission calculations by including specific recombination rates for H~I, O~I, Mg~I, Na~I, Fe~I and Fe~II\footnote{With the total radiative recombination rates taken from \href{www.astronomy.ohio-state.edu/~nahar/nahar_radiativeatomicdata/index.html}{www.astronomy. ohio-state.edu/$\sim$nahar/nahar\_radiativeatomicdata/index.html}, as in \texttt{SUMO}}. For the other elements, we follow the default treatment as in \texttt{SUMO}, allocating 1/3rd of the rate into the ground state multiplet and the rest distributed following statistical weights. Free-bound emission is calculated using the following (see also Eq~B.22 in \citet{jerkstrand2012progenitor}):
\begin{equation}
    j^{\text{f-b}}_\text{k,l} = \frac{1}{4\pi}n_e n_+ \alpha_\text{k,l}(T) \ \times\  \Big( 0.8kT + \chi_\text{k,l}\Big),
    \label{eq:freebound_emiss}
\end{equation}
which is emitted over a box-shaped profile between $0.5-1.5\,$kT. $n_e$ is the free electron density, $n_+$ is the number density of the ion, $\alpha_\text{k,l}(T)$ is the specific recombination coefficient and $\chi_\text{k,l}$ is the energy difference between the starting level and ionization threshold to reach this ion.

The cross section $\sigma_\text{PI}$ due to this approximate treatment is calculated using the hydrogenic approximation \citep{rybicki1979radiative} as follows (see also \citealt{jerkstrand2012progenitor}, appendix B6):
\begin{equation}
    \sigma_\text{PI} = 7.9\times10^{-18} \times \bigg(\dfrac{\lambda_\text{bin}}{\lambda_\text{ion}}\bigg)^3 \times \dfrac{\Big(1-\dfrac{E_u}{E_\text{ion}}\Big)^{-0.5}}{(\rm{C_\text{ion}+1)^2}}~\rm{cm}^2,
    \label{eq:sigma_OTSPI}
\end{equation}
where $\lambda_\text{bin}$ is the wavelength of the bin where the emission is emitted, $\lambda_\text{ion}$ is the ionization threshold up to which a photon can ionize, $E_u$ is the energy of the starting level, $E_\text{ion}$ is the ionization threshold, C$_\text{ion}$ is the charge of the starting level, and where the Gaunt factor has been assumed to be unity. The optical depth $\tau_\text{PI}$ is then determined as follows:
\begin{equation}
    \tau_\text{OTS-PI} = \sigma_\text{PI} \times n_u \times V^{1/3}_\text{cell},
    \label{eq:tau_OTSPI}
\end{equation}
where $n_u$ is the level population of this level, and $V_\text{cell}$ is the volume of the cell where the recombination takes place. Taking $V^{1/3}$ is a rough proxy for a length scale that the photons would travel in each cell, which might overestimate the absorption in the cell somewhat. However, since the approximation is only applied in the local cell, this partially evens out to some of this emission escaping to the next cell and being absorbed there which would happen in a non-local case. 
The absorbed emission is accounted for in the \textsc{T\_solver} as an additional heating source, and in the \textsc{NLTE\_solver} as photoionization.

All recombination emission is binned into bins with a width of $50\,\angstrom$. First all $\tau_\text{OTS-PI}$ rates are calculated and then the actual rates are assigned depending on the relative contribution from each level, from each ion, to the total $\Sigma\tau_\text{OTS-PI}$ within the bin where the emission took place. The binning is relatively rough, as the calculation of $\sigma_\text{PI}$ is relatively expensive compared to the rest of the \textsc{NLTE\_solver}. This specific treatment can still be used by \texttt{ExTraSS} in the first iteration of the \textsc{NLTE\_solver}, as a replacement for the not-yet computed radiation field.

\section{MPI communication bundle sizes}   \label{app:bundletesting}
In this Appendix, we list some small, older tests performed to find a robust and optimal size for the communication bundles within the domain decomposition scheme, both between `managers' and `workers' intra-node as well as between two managers inter-node, for the \textsc{RayTraceGrid} module. The scheme itself is described in Section~\ref{sssec:dodec_method}.

Several different sizes for both the worker-to-manager bundle size as well as the node multiplier were tested, with an overview given in Table~\ref{tab:bundle_sizes_data}. They were tested on the Dardel system (see main text), 4 nodes with 128 cores each.

\begin{table}
\centering
\begin{tabular}{c c c c c c}
\hline
Bundle Size & Multiplier & $t_\text{last cell}$ & $t_\text{term}$ & $\delta$t & Ranking \\
  &  & (s) & (s) & (s) & $t_\text{lc}$ : $t_\text{term}$ \\
\hline
16384  & 18x & 9057 & 9412 &  355  & 10 :  9  \\
32768  & 12x & 8159 & 8746 &  587  &  2 :  7  \\
32768  & 24x & 8252 & 8296 &   44  &  4 :  2  \\
49152  & 24x & 8284 & 8504 &  220  &  5 :  4  \\
57344  & 24x & 9034 & 9473 &  439  &  9 : 10  \\
65536  & 20x & 8294 & 8665 &  371  &  6 :  6  \\
65536  & 22x & 8133 & 8839 &  706  &  1 :  8  \\
65536  & 24x & 8241 & 8289 &   48  &  3 :  1  \\
65536  & 26x & 8351 & 8564 &  213  &  8 :  5  \\
65536  & 30x & 8309 & 8349 &   40  &  7 :  3  \\
\hline
\end{tabular}
\caption{Performance metrics, using $3\times10^9$ rays across 20\,000 cells with 512 cores across 4 nodes (the setup as shown in Figures~\ref{fig:DoDec_sketch} and \ref{fig:MPIcomm}), for worker-to-manager bundle size and node multiplier. The $t_\text{last cell}$ indicates how quickly the program finished all cells, while the $t_\text{term}$ indicates how long it took the program to finish up with all the packets that were still `stuck' in the buffers at $t_\text{last cell}$. The time for this `flushing' ($\delta$t) is also listed. At different bundle sizes there are different optimal multipliers.}
\label{tab:bundle_sizes_data}
\end{table}

As shown by the timing differences in Table~\ref{tab:bundle_sizes_data}, there is no monotonic relation between the bundle size and multiplier together. Generally, too small of a bundle leads to too much communication and this drags down the efficiency, while having too small of a multiplier instigates too many (smaller) communication streams between the different nodes. Several setups achieved very similar outcomes; however, a full 3D grid might send around a factor 100 more rays than used in the trial runs, and when more total rays are send, larger bundle sizes become preferred. As such, we opted for 65536 for the current setup, but this number is flexible. We tested several different node multipliers to accompany this bundle size, and settled on 24x as this was the fastest overall by a small margin, while also having one of the smallest $\delta$t `flushing' times.

\section{\txtred{Level population details}}   \label{app:levelpop_compare}
\txtred{In this Appendix, we take a look at the level populations for the 1D model of \J18 (with \texttt{SUMO}) and the 3D version thereof computed with \texttt{ExTraSS}. Spectrally, these two models are very similar (see Figure~\ref{fig:J18_3Dcomp}) although the rates for photoionization (Figure~\ref{fig:PIrates_1D3D}) and photoexcitation (Figure~\ref{fig:PErates_1D3D}) can still differ by an order of magnitude at certain points between the codes. We will first look at how the ionization rates impact the neutral fractions of O and Fe throughout the \J18 model alongside a comparison to the 3D explosion model from \vBPIII26. Afterwards, we will focus on the level populations in O I and Ca II, as oxygen and calcium are important emitters in the nebular phase, and these are the key ions for these species, and also the full level population structure of Fe, as it is an important species in the \J18 model.}

\subsection{\txtred{Neutral fractions}}
\begin{figure}
    \centering
    \includegraphics[width=\linewidth]{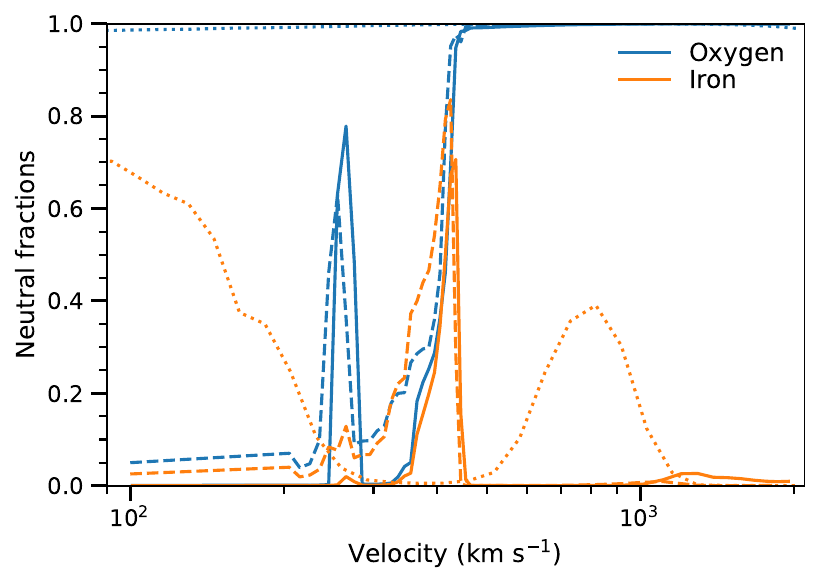}
    \caption{A comparison between the neutral fractions of oxygen (blue) and iron (orange), for the same three models as in Figure~\ref{fig:PIrates_1D3D}, with \texttt{SUMO} \txtred{values} from \J18 in dashed, and \texttt{ExTraSS} \txtred{values} from that model in 3D in solid, and from \vBPIII26 in dotted lines. The 3D \txtred{values} shown are the median fractions.}
    \label{fig:ionfracs_FeO}
\end{figure}
In Figure~\ref{fig:ionfracs_FeO} we show how the different \txtred{photo}ionization rates between the models translate to ionization stage fractions for oxygen (in blue) and iron (orange) -- we do not show calcium as the neutral Ca fraction is always negligible. The fractions for the model from \J18 are largely similar for 1D and 3D, although some minor differences appear. For the innermost ejecta (\kms{v<250}), where the photoionization rates were different by \txtred{roughly} an order of magnitude, \texttt{SUMO} finds neutral fractions of up to 0.1 for oxygen and 0.05 for iron, while in \texttt{ExTraSS} with the higher photoionization rates these species are completely ionized. At \kms{v\approx250} there is a dip in the photoionization rates, and correspondingly the neutral fractions for particularly oxygen increase sharply, to $\sim0.6$ for \texttt{SUMO} and $\sim0.8$ for \texttt{ExTraSS}. Between \kms{300\text{ to }450} the photoionization rates are more similar for both codes, and also the neutral fractions agree quite well, particularly for oxygen. \txtred{Past \kms{1000}, there is a small fraction of neutral iron found in \texttt{ExTraSS}, which is not present in \texttt{SUMO}.}

The structure of the 3D explosion model from \vBPIII26 is completely different, and as such the neutral fractions (dotted lines in Figure~\ref{fig:ionfracs_FeO}) for this model do not follow the same trends at all. Oxygen has a much stronger presence in the innermost \kms{400} (at mass fractions of $\sim0.05-0.1$) for this model, which also has a higher density. The photoionization rate for O~I is also much lower at these low velocities, and thus oxygen is predominantly neutral in this model. Iron acts differently, as the innermost \kms{v\approx250} are mixed between envelope and Fe-core material. As such, the photoionization rates for Fe are much lower in \vBPIII26 (see Figure~\ref{fig:PIrates_1D3D}), making Fe relatively neutral at the lowest velocities. Towards the outer envelope, the photoionization rates drop again, leading to a rising neutral fraction up to $\sim0.4$ at \kms{v\approx800}. At these velocities, part of the \vBPIII26 model has little $\gamma$-ray energy deposition, making it much cooler and more neutral.

The difference in the photoionization rates \txtred{(see Figure~\ref{fig:PIrates_1D3D})} for Ca between the models is smaller than for O and Fe, two elements which still obtain largely similar ionization structures for the \J18 model. In both setups, Ca is at least singly ionized. 

\subsection{\txtred{Level population structure}}
\begin{figure*}
    \centering
    \includegraphics[width=.8\linewidth]{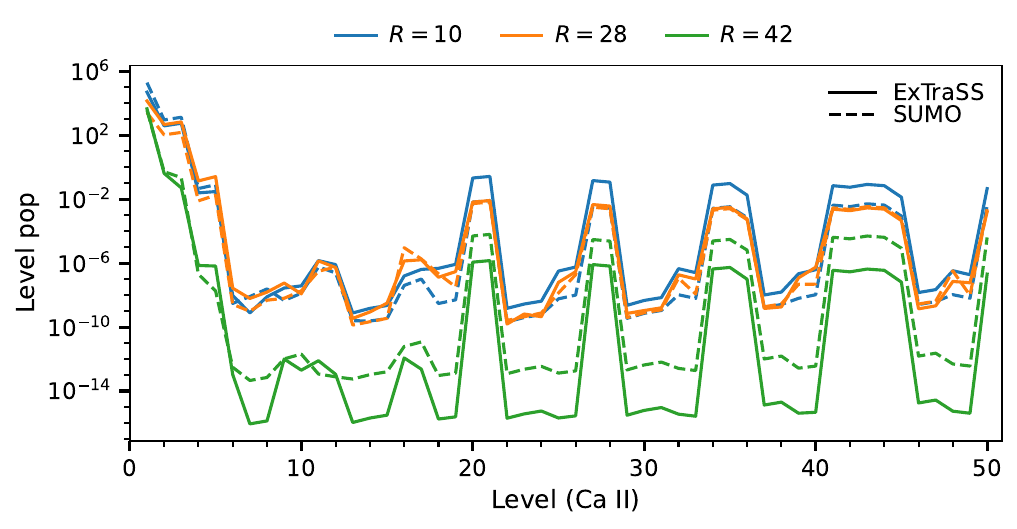}
    \includegraphics[width=.8\linewidth]{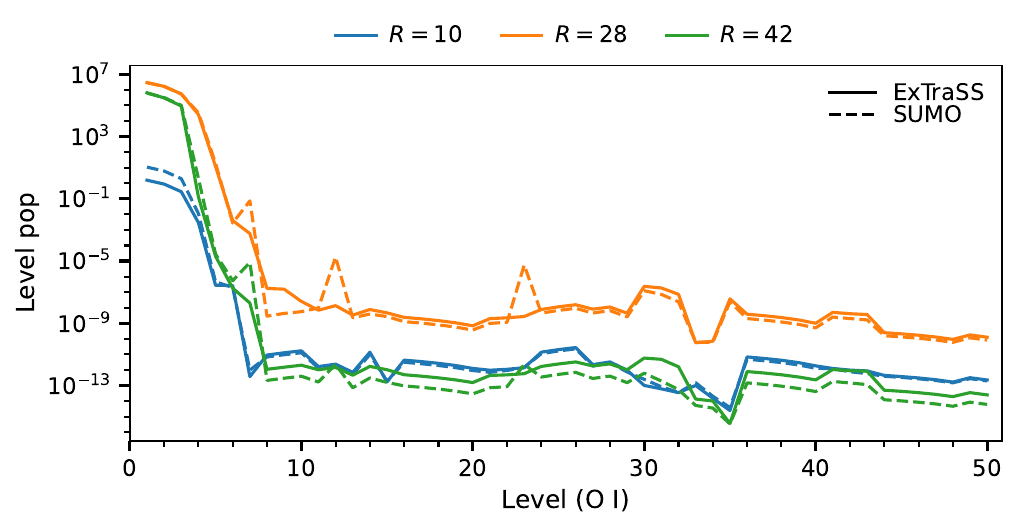}
    \caption{The level populations for Ca~II (top) and O~I (bottom) for the 1D model from \J18 (with \texttt{SUMO}, dashed lines) and with \texttt{ExTraSS} (solid lines), using the 1D-turned-3D model. The \txtred{radii shown are $R=10$ at \kms{285}, $R=28$ at \kms{465} (right after the `wall' between the Ni-rich core material, inside the He-envelope) and $R=42$ at \kms{1500}. For readability,} only the first 50 levels are shown.}
    \label{fig:CaO_levelpopts}
\end{figure*}
\begin{figure*}
    \centering
    \includegraphics[width=.8\linewidth]{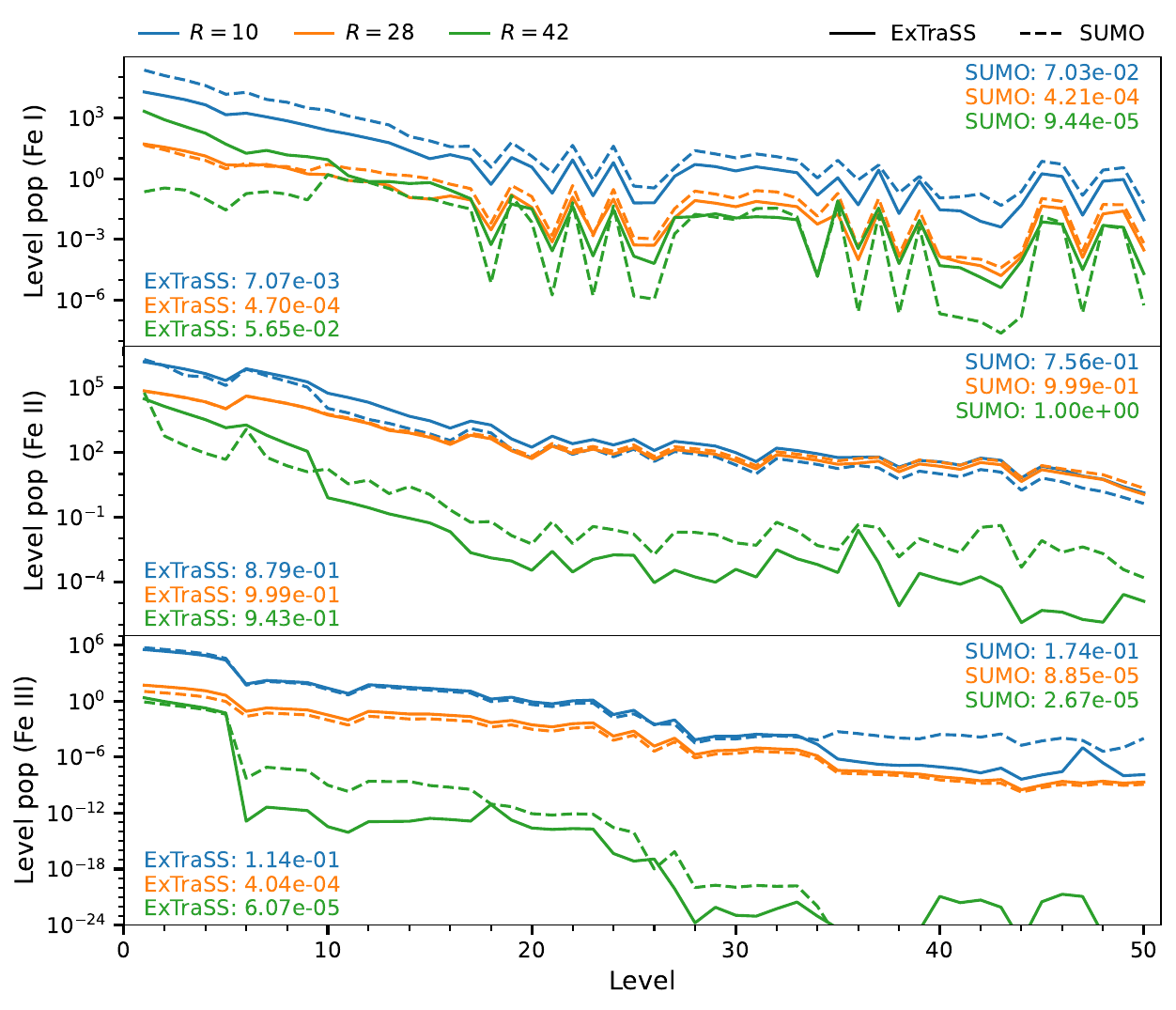}
    \caption{Comparison of the Fe level populations for the 1D model from \J18 (\texttt{SUMO}, dashed lines) against the level populations from the 3D version of this model computed by \texttt{ExTraSS} (solid lines), at \txtred{the same three radii as Figure~\ref{fig:CaO_levelpopts}}. The top panel shows \txtred{Fe~I, the middle panel Fe~II, and the bottom panel Fe~III}. The plots are truncated at level 50 for readability. \txtred{The ion fractions are shown in the top right (\texttt{SUMO}) and bottom left (\texttt{ExTraSS}) of each panel.}}
    \label{fig:Fe-SUMO:ExTraSS}
\end{figure*}
In Figure~\ref{fig:CaO_levelpopts} the level populations of Ca~II (top panel) and O~I (bottom panel) are shown (for readability, the plots are truncated at level 50) in the \J18 model, for \texttt{ExTraSS} (solid) and \texttt{SUMO} (dashed). We showcase three different radii, at various points in the ejecta. For $R=10,\,28$ the agreement between \texttt{ExTraSS} and \texttt{SUMO} is very good for Ca~II, and at each of the radii for O~I, although some differences appear in the level populations between the two codes. For Ca~II, \texttt{ExTraSS} \txtred{finds that at $R=10$, the lowest levels are slightly less populated and the higher laying states are somewhat more populated but still at subdominant fractions. At $R=42$, \texttt{ExTraSS} finds lower level populations for all levels past the first 6, although these levels have marginal population for both codes at this radius.} For O~I, in particular at $R=28$, there are a few levels in \texttt{SUMO} which have a large spike in their population, while in \texttt{ExTraSS} these do not occur. \txtred{At $R=10$, the fraction of O I is quite low for both codes (see Figure~\ref{fig:ionfracs_FeO}), but somewhat lower for \texttt{ExTraSS} still, which explains the difference in the ground states.}

Overall, \txtred{for these ion species}, \texttt{SUMO} and \texttt{ExTraSS} find largely the same structure for the level populations, although differences appear in particular at higher velocities. When accounting for the different ionization structures, the differences do become smaller.

In Figure~\ref{fig:Fe-SUMO:ExTraSS} we look at the level populations \txtred{for the first 50 levels of the first three ionization stages of Fe (Fe~I, Fe~II and Fe~III, from top to bottom)} in the \J18 model, for \texttt{ExTraSS} (solid) and \texttt{SUMO} (dashed)\txtred{, for the same three radii as before. Unlike O and Ca, Fe is not dominated by one ion species across all radii, so all three are shown together. In the top right of each panel, the ion fractions for \texttt{SUMO} are shown at each radius, and the \texttt{ExTraSS} fractions are shown in the bottom left.} Generally, there is quite good agreement between the codes, although some differences appear. \txtred{At $R=10$, \texttt{SUMO} is less dominated by Fe~II (0.879 for \texttt{ExTraSS}, versus 0.756 for \texttt{SUMO}) which results in higher level populations in both Fe~I and Fe~III in \texttt{SUMO}. At $R=28$, Fe~II is the dominant ion for both, and the codes find good agreement although \texttt{ExTraSS} finds somewhat lower level populations in Fe~I past level 10. At $R=42$, \texttt{SUMO} is still completely dominated by Fe~II, while \texttt{ExTraSS} is only at 0.943, which results in quite a noticeable difference in Fe~I.}

\txtred{Some of this offset in this last radius can be connected back to the spectral differences in Figure~\ref{fig:J18_3Dcomp}, where \texttt{ExTraSS} has marginally lower Fe I emission in the range $7900-8400$ as well as less emission at $\sim4200$ and the blended $5180\,\angstrom$ Mg+Fe feature. This might seem unexpected, as the Fe~I fraction in \texttt{ExTraSS} is noticeably higher at the outer radii, but as Figure~\ref{fig:Fe-SUMO:ExTraSS} (top panel, green curves) shows, most of this difference presents itself in the lowest 10 levels, and the Fe emission in these regions comes from transitions towards those levels. With the higher level populations, the Sobolev optical depth (see Equation~\ref{eq:tau_Sobolev}) will change noticeably and thereby alter the emission from the outer radii.}

\txtred{The stark difference in level populations for Fe I at the outer radii might be caused by runaway ionization \citep[see][Appendix E]{jerkstrand2017long}, which would cause the strong drop in level populations for the lowest levels, while also strongly increasing the level population of the ground state of Fe II as the target state for the ionizations. Having a depleted Fe I population would also lead to the high photoionization rate for Fe I observed past \kms{1000} (Figure~\ref{fig:PIrates_1D3D}). The higher photoionization rates of \texttt{ExTraSS} at lower velocities does not seem to lead to runaway ionization, as the level population curves at $R=10$ follow the same pattern between both codes, which suggests that while there is more Fe II, the \textsc{NLTE\_solver} is not pushed into catastrophic depletion of certain states.}


\bsp	
\label{lastpage}
\end{document}